\documentclass[aps,showpacs,preprintnumbers,amsmath,amssymb, 
twocolumn, tightenlines]{revtex4}

\usepackage[dvips]{graphicx}\usepackage{bm}
\usepackage{amsmath}
\usepackage{dcolumn}
\usepackage{bm}

\newcommand{\be}{\begin{equation}}
\newcommand{\ee}{\end{equation}}
\newcommand{\ba}{\begin{eqnarray}}
\newcommand{\ea}{\end{eqnarray}}
\newcommand{\no}{\nonumber \\}
\newcommand{\gsim}{\mathrel{\hbox{\rlap{\lower.55ex \hbox {$\sim$}}
                   \kern-.3em \raise.4ex \hbox{$>$}}}}
\newcommand{\lsim}{\mathrel{\hbox{\rlap{\lower.55ex \hbox {$\sim$}}
                   \kern-.3em \raise.4ex \hbox{$<$}}}}
\def\be{\begin{eqnarray}}
\def\ee{\end{eqnarray}}

\def\del{\partial}

\def\vk{{\vec k}}

\def\roughly#1{\mathrel{\raise.3ex\hbox{$#1$\kern-.75em%
\lower1ex\hbox{$\sim$}}}}
\def\lsim{\roughly<}
\def\gsim{\roughly>}

\def\vx{{\vec x}}

\def\({\left(}
\def\){\right)}
\def\[{\left[}
\def\]{\right]}

\def\N{${\cal N}\,\,$}

\def\I{\bigcirc\hspace{-0.25cm}1}
\def\II{\bigcirc\hspace{-0.25cm}2}
\def\III{\bigcirc\hspace{-0.25cm}3}

\def\ka{{\kappa}}
\def\lam{{\lambda}}
\def\dlt{{\delta}}
\def\omg{{\omega}}
\def\tlam{{\tilde{\lambda}}}
\def\tlt{{\tilde{t}}}
\def\tlr{{\tilde{r}}}
\def\tlx{{\tilde{x}}}
\def\tlp{{\tilde{P}}}
\def\tp{{\tau_+}}
\def\tm{{\tau_-}}
\def\lv{\lvert}
\def\rv{\rvert}
\def\gm{{\gamma}}
\def\bt{{\beta}}

\begin{document}
\title{Toward the AdS/CFT Gravity Dual for High Energy Collisions:
\\ II.
  The Stress Tensor on the Boundary   }

\author{Shu Lin and Edward Shuryak}
\affiliation{ 
Department of Physics and Astronomy,  Stony Brook University , 
Stony Brook NY 11794-3800, USA
}

\date{\today}
\begin{abstract}
In this second paper of the series we calculate the stress tensor
of excited matter, created by ``debris'' of high energy collisions
at the boundary. We found that massive objects (``stones'') falling 
into the AdS center produce gravitational disturbance which however 
has zero stress tensor at the boundary
The falling open strings, connected to
receeding charges, produce a nonzero stress tensor which we
found analytically from
time-dependent linearized Einstein equations in the bulk.
It corresponds to exploding non-equilibrium matter: we discuss
its behavior in some detail, including its internal energy
density in a comoving frame and the ``freezeout surfaces''.
We then discuss what happens for the ensemble of strings. 
\end{abstract}
\maketitle
\vspace{0.1in}

\section{Introduction}

This is the second paper of the series, and the first
one \cite{Lin:2006rf} (to be referred to as [I] below) had rather extensive introduction.
Therefore
we just  briefly reiterate here what are the main goals of this study.

Holographic description of \N=4 SYM theory in strong coupling
regime can be achieved via AdS/CFT correspondence, which
relates it to string theory in $AdS_5\times S^5$ space,
in classical supergravity regime. 
Large number of applications use this tool to study
properties of strongly coupled Quark-Gluon plasma: but most of them
are done in a static setting, with fixed temperature via Witten's
AdS black hole metric.

High energy hadronic collisions in QCD are very difficult problems.
They are time-dependent and include non-equilibrium physics
(only collisions of heavy ions can be approximated by hydrodynamics
and locally equilibrated QGP). On top of that
 they involve different scales and different coupling 
regimes. Pure phenomenological approaches were developed long ago,
such as e.g. Lund model \cite{Andersson:1983ia} which are based
on a picture of QCD strings stretched by departing partons.
More recent approach -- known as  the Color Glass
  picture -- was proposed by McLerran and Venugopalan
\cite{MLV} who argued that
since fluctuations
 high energy collisions lead to large local color charges
(in the transverse plane) they would lead to
 production of strong color fields. Those are
 treated
by classical Yang-Mills eqn in a {\em weak coupling} regime.

   Arguments suggested recently put forward a view
 that QCD has certain  ``strong coupling window''. 
 In particular, Brodsky and Teramond \cite{Brodsky:2007vk} 
have argued that
 the power  scaling observed for large  number of exclusive processes
 is not due to perturbative QCD (as suggested originally in 1970's) but
 to a  strong coupling regime in which the running is absent
and quasi-comformal regime sets in.  Polchinski and Strassler
 \cite{Polchinski:2001tt} have shown that in spite of exponential
string amplitudes one does get power laws scaling for exclusive
processes, due to convolution (integration over the $z$ variable)
with the power tails of hadronic wave functions.
 One of us
 proposed a  scenario \cite{mydomain} for AdS/QCD
 in which there are two domains, with weak and strong coupling. The
 gauge coupling rapidly rises at the ``domain wall''
associated with instantons. Pion diffractive dissociation
is a process where a switch and weak coupling domains are observed:
and cross section behavior is consistent with Polchinski-Strassler
expression and expected coupling change.
 Last but not least,
 such approach looks now natural in comparison to what happens in
 heavy ion/finite T QCD, where we do know now that at comparable parton
 densities  the system indeed is in a strong coupling regime. 

Accepting the Color Glass
  picture as an asymptotic  for very high parton density
 and large saturation scale $Q_s\rightarrow \infty$, one
should  ask what should happen
 in the case of saturation scale falling to intermediate momenta
  $Q_s\sim .3-1.5 \, GeV$ associated with ``strongly coupled window''.
 This is the issue we address in this work,
 using the AdS/CFT correspondence in its time-dependent version,
as a tool to describe the evolution of the system.

 The setting has been discussed in [I], where we 
extensively studied how exactly
the ``debris'' produced in a collision --
 particles or open strings -- are
falling under gravity force into IR (the AdS
 throat). In this work
we do the next technical step
and calculate the back reaction of gravity  by solving the linearized
Einstein eqns for metric pertubations, deducing the
space-time dependence
of the stress tensor  $T_{\mu\nu}(x)$ of excited matter observed on the boundary.

 The general setting is in fact rather similar
 to the Lund model: 
except that strings  are departing from
our world ($z=0$ boundary) rather than breaking.  
Technically our work is a development along
a line actively pursued by many authors. 
(Early work in a different scenario focused on the effective stress tensor on a brane\cite{Giddings:2000ay}).
In particular,
it can be considered a continuation of our recent work 
\cite{Lin:2007pv} in which we calculated static (time-independent)
stress tensor associated with Maldacena's static dipole.
It is also similar to recent AdS/CFT calculations 
of a hydrodynamical
``conical flow'' from quenching jet in QGP 
 \cite{Gubser,Chesler:2007an}.

The process we describe  resembles what happens in 
heavy ion collisions,
but with very important distiction. In the setting of this paper
we treat ``debris'' as small pertubation, solving linearized
Einstein eqns in pure $AdS_5$ background.
 Therefore there is no black hole and/or temperature
in this work, and our ``mini-explosion'' produce
 matter which is not equilibrated and the resulting
stress tensor cannot be parameterized  hydrodynamically.

To get all that one needs to proceed to nonlinear gravity and
a non-linear process of  black hole formation: the
problem which we hope to attack elsewhere.

\section{Solving Linearized Einstein equation}

We want to solve linearized Einstein equation in 
 $AdS_5$ background in Poincare
 coordinates, with standard background  metrics
\be
ds^2=\frac{-dt^2+d\vx^2+dz^2}{z^2}+h_{\mu\nu}
\ee
An axial gauge is chosen for
 metric perturbation, so $h_{z\mu}=0$

The linearized Einstein equation are well known
and the tactics used in its solution are 
discussed in  \cite{Lin:2007pv}: the present case is
only different by appearence of time derivatives.
We put it into the form

\be\label{eq:hmn}
{1\over 2}\square h_{mn}-2h_{mn}+{z\over 2}h_{mn,z}=s_{mn}
\ee
where $\square=z^2\(-\del_t^2+\del_{\vx}^2+\del_z^2\)$, 
the indices are 0-3 and the r.h.s. is the generalized source 
\be &&s_{mn}=\dlt S_{mn}-\int_0^z\(\dlt S_{zm,n}+\dlt S_{zn,m}\)dz
\no
&&+{1\over 2}h_{,m,n}+{1\over 2}\Gamma_{mn}^z h_{,z} \ee 
containing not only the stress tensor of the source
\be \dlt S_{\mu\nu}=-\ka^2(T_{\mu\nu}-\frac{T}{3}g_{\mu\nu}) \ee
but also the following combinations of perturbation metric
which can be easily found from the eqns:
\be\label{eq:h}
&&h={1\over 3}\int_0^z dz\cdot z
\biggl[
\dlt S_{zz}+\dlt S_{tt}-\Sigma_i \dlt S_{x^ix^i}+ \no
&&2\int_0^z dz \(-\dlt S_{zt,t}+\Sigma_i \dlt S_{zx^i,x^i}\)
\biggr] \nonumber
\ee



The source term for different objects
 is  standard, obtained via variation of their action
 over the metric
\be\label{eq:Tmn}
&&S_{NG}=-\frac{1}{2\pi\alpha'}\int d^2\sigma\sqrt{-det g_{ind}}
\int d^5x\dlt^{\(5\)}\(x-X\(\sigma\)\)      \no
&&T^{\mu\nu}=-\frac{2\dlt S_{NG}}{\sqrt{-g}\dlt g_{\mu\nu}} \no
&&=\frac{1}{\sqrt{-g}2\pi\alpha'}\int d^2\sigma\dlt^{\(5\)}\(x-X\(\sigma\)\)
\del_{\alpha}X^{\mu}\del_{\beta}X^{\nu}g^{\beta\alpha}_{ind} \\
&&S_m=m\int d^5x\dlt^{(5)}(x-X(s))\int ds\sqrt{-g_{\mu\nu}\frac{dx^{\mu}}{ds}\frac{dx^{\nu}}{ds}} \no
&&T^{\mu\nu}=-\frac{m}{\sqrt{-g}}\int ds\dlt^{(5)}(x-X(s))\frac{\frac{dx^{\mu}}{ds}\frac{dx^{\nu}}{ds}}{\sqrt{-g_{\mu\nu}\frac{dx^{\mu}}{ds}\frac{dx^{\nu}}{ds}}}
\ee
here  $g_{\mu\nu}$ and $g_{ind,\alpha\beta}$  denote the AdS metric
and the induced metric on the string worldsheet,
 respectively.

In the next section we find an expression for Green's function,
which will provide $h_{mn}$ for
any given source $s_{mn}$. We will then exctract an expression
for the coefficient of the $z^2$ term in Taylor series 
of $h_{mn}$ at the boundary, which by the rules of AdS/CFT
correspondence gives
us the boundary stress tensor.

\section{The Green's function for the linearized gravity in $AdS_5$}

The Green's function we need
satisfies the following eqn:
\be\label{eq:green_x}
&&\frac{z^2}{2}(\del_z^2-\del_t^2+\del_{\vx}^2)G(z,z')-2G(z,z')+
\frac{z}{2}\del_z G(z,z') \no
&&=\dlt(z-z')\dlt(t-t')\dlt^{(3)}(x-x')
\ee
and the solution to (\ref{eq:hmn}) is then given by
 $h_{mn}(z)=\int G(z,z')s_{mn}(z')dz'$. Thus $G(z,z')$ should
 satisfy the same boundary condition as $h(z)$.
Fourier transforming 4-dim part of (\ref{eq:green_x}), we have $z$-
dependent eqn
\be\label{eq:green_k}
&&\frac{z^2}{2}(\del_z^2+\omg^2-k^2)G(z,z')-2G(z,z')+
\frac{z}{2}\del_z G(z,z') \no
&&=\dlt(z-z')
\ee
where  $G^k(z,z')=\int G(z,z')e^{-i\omg t+i\vk\vx}dtd^3x$

(\ref{eq:green_k}) can be solved in terms of Bessel functions:
For $\lv\omg\rv>k$, the solution is a linear combination
of Bessel functions of the first and second kind. We impose
the following boundary condition: at $z=0$,
$G(z,z')=0(h(z)=0)$, at $z\rightarrow \infty$, $G(z,z')(h(z))$
contains outgoing wave only, i.e. the wave propagates from the
 source to infinity\footnote{the same boundary condition is used
 in \cite{son:correlator} as a limiting case of thermal AdS background}.
The solution is composed of two homogeneous solutions:

\be\label{eq:G1}
G(z,z')=\lbrace
\begin{array}{ll}
A J_2(\lam z)& z<z'\\
B(J_2(\lam z)-isgn(\omg)Y_2(\lam z))&z>z'
\end{array}
\ee
with $\lam=\sqrt{\omg^2-k^2}$, $A,B$ is fixed by matching the
function itself and its first derivative at $z=z'$:

\be
\lbrace
\begin{array}{ll}
A=\frac{\pi isgn(\omg)}{z'}(J_2(\lam z')-isgn(\omg)Y_2(\lam z'))\\
B=\frac{\pi isgn(\omg)}{z'}J_2(\lam z')
\end{array}
\ee

For $k>\lv\omg\rv$, the solution can be built from Modified 
Bessel function. We choose the  no exponential growth
boundary condition at $z\rightarrow \infty$\cite{Lin:2007pv}. The solution
is given by:

\be\label{eq:G2}
G(z,z')=\lbrace
\begin{array}{ll}
C I_2(\tlam z)& z<z'\\
D K_2(\tlam z)& z>z'
\end{array}
\ee
with $\tlam=\sqrt{k^2-\omg^2}$
\be
\lbrace
\begin{array}{ll}
C=\frac{-2K_2(\tlam z')}{z'}\\
D=\frac{-2I_2(\tlam z')}{z'}
\end{array}
\ee

It turns out the solution can be organized in a compact form 
using properties of Bessel function:
\be
G(z,z')=\lbrace
\begin{array}{ll}
-{2\over{z'}}I_2(i\lam z_<)K_2(i\lam z_>)& \omg>0,\lv\omg\rv>k\\
-{2\over{z'}}I_2(-i\lam z_<)K_2(-i\lam z_>)& \omg<0,\lv\omg\rv>k\\
-{2\over{z'}}I_2(\tlam z_<)K_2(\tlam z_>)& \lv\omg\rv<k
\end{array}
\ee
where $z_<=min(z,z')\;z_>=max(z,z')$

Doing the inverse Fourier transform: 
$\frac{1}{(2\pi)^4}\int G^k(z,z')e^{i\omg t-i\vk\vx}d\omg d^3k$, 
we obtain a retarded propagator for the metric:
(See Appendix A for the evaluation of the integral, a retarded scalar 
propagator was found in \cite{vacua_probes}. Other scalar and graviton 
propagators 
were also found in \cite{Giddings:2000mu} with slightly different boundary conditions)
\be
&&P_R=\frac{12iz}{(2\pi)^2}
\biggl[\frac{1}{(t^2-r^2-z^2+i\epsilon)^4}- \no
&&\frac{1}{(t^2-r^2-z^2-i\epsilon)^4}\biggr]
\theta(t-r)
\ee

Several comments of the propagator are in order:
(i)The theta function implies the propagator acts inside the lightcone
$t^2-r^2=z^2>0$ and moreover is retarded $t>r>0$, which we indicate by the subscript R. It is also
consistent with the outgoing boundary condition. Note that
 the propagator
is also Lorentz invariant.
(ii)The propagator relates the $z^2$ coefficient of metric perturbation $Q_{mn}$
 and the source $s_{mn}$ in the following way(assuming $s_{mn}(z)$
 does not contain $z^0$ and $z^2$ terms):
$Q_{mn}(t',x')=\int P_R(t'-t,x'-x,z)s_{mn}(z,t,x)dzdtd^3x$
The primed and unprimed coordinates correspond to boundary and bulk
respectively.
(iii)The propagator is dynamical. For static source, one can 
perform the t-integral to obtain a static propagator, which agrees with
the one obtained in \cite{Lin:2007pv}.

\section{The stress tensor of a falling stone is zero!}

First we want to study the stress tensor induced by a falling stone. The 
trajectory of stone in $AdS$ background is: $z(t)=\sqrt{z_m^2+t^2}\equiv 
{\bar z}$ for $t>0$. Plugging the trajectory, it is not difficult to obtain the generalized
source:

\be\label{eq:gstone}
&&s_{mn}=\frac{\ka^2m\dlt^{(3)}(\vx)}{3z_m}
\biggl[
\begin{pmatrix}
2& \\
& 1
\end{pmatrix}{\bar z}^3\dlt(z-{\bar z})+
\begin{pmatrix}
1& \\
& -1
\end{pmatrix}\times \no
&&{\bar z}t^2\dlt(z-{\bar z})+
\begin{pmatrix}
2\del_t& \del_{x_{i}}\\
\del_{x_{i}}& 
\end{pmatrix}3{\bar z}^2t\theta(z-{\bar z})+
\begin{pmatrix}
\del_t^2& \del_t\del_{x_{i}}\\
\del_t\del_{x_{i}}& \del_{x_{i}}\del_{x_{j}}
\end{pmatrix}\times \no
&&({\bar z}^2+2t^2)\theta(z-{\bar z})+
\begin{pmatrix}
-1& \\
& 1
\end{pmatrix}({\bar z}^2+2t^2)\theta(z-{\bar z})
\biggr]
\ee

Convolute the generalized source with the propagator $P_R$. We use the 
substitution: ${\del_{x_m}}=-{{\overleftarrow\del}_{x_m}}={{\overleftarrow\del}_{x_m'}}$ (where $x_m=t,x^i$) to simplify the calculation. We use 
partial integration in the first identity so that the derivative acts on the
propagator, which we indicate by an over left arrow. The second identity
is due to the fact $P_R=P_R(x_m'-x_m)$. After that we can see that the $x^i$
 integral kills the $\dlt^{(3)}(\vx)$ and the $z$-integral involving
either delta function or theta function can also be done easily. Finally 
we are left with the $t$-integral:

\be
&&Q_{mn}=\frac{\ka^2m}{3z_m}\int dt
\biggl[
\begin{pmatrix}
2& \\
& 1
\end{pmatrix}{\bar z}^4\tlp+
\begin{pmatrix}
1& \\
& -1
\end{pmatrix}{\bar z}^2t^2\tlp+ \no
&&\begin{pmatrix}
2(t'-t)& -x_i'\\
-x_i'& 
\end{pmatrix}3{\bar z}^2t\tlp+
\begin{pmatrix}
(t'-t)^2& -(t'-t)x_i'\\
-(t'-t)x_i'& x_i'x_j'
\end{pmatrix}\times \no
&&({\bar z}^2+2t^2)\tlp
\biggr]
\ee
with
\be
&&\tlp=\frac{12i}{(2\pi)^2}
\biggl[\frac{1}{((t'-t)^2-r'^2-{\bar z}^2+i\epsilon)^4}- \no
&&\frac{1}{((t'-t)^2-r'^2-{\bar z}^2-i\epsilon)^4}\biggr]\theta(t'-t)
\ee

The $t$-integral is evaluated via residue theorem. Notice $\tlp$ contains
a fourth order pole at $t=\frac{t'^2-r'^2-z_m^2}{2t'}$
in the denominator, only the coeffcients of $t^3$ and $t^4$
in the overall numerator are relevant for the integral. Summing up the
matrices, we end up with a vanishing result.\footnote{We have assumed that 
the pole lies in the integration range of $t$ in using the residue theorem.
In fact if it is not the case, we will still get a vanishing result due to
the simple cancellation between $\pm$ term in $\tlp$}. It is probably not
a coincidence that $trF^2$\cite{vacua_probes}
calculated from dilaton perturbation vanished as
well. Both calculations are in linearized approximation and correspond to
 a falling stone. Nevertheless the nonlinear version of stress tensor 
induced by the same object  obtained in \cite{shock,stone} seems to contradict 
our result, which deserves further study.

\section{The stress tensor of a falling open string}

We want to study the stress tensor by a falling string.
A scaling string profile is obtained in [I] for a separating quark-
antiquark pair, provided the separating velocity is not too large: $v<0.6$.
We briefly recall the scaling solution. The quark(antiquark)
moves along the trajectory $x=\pm vt$. The string profile is given by:

\be\label{eq:scaling}
&&z=\frac{\tau}{f(y)} \no
&&y=f_0\sqrt{\frac{f_0^2-1}{2f_0^2-1}}
F\(\sqrt{\frac{f^2-f_0^2}{f^2-1}},\frac{f_0}{\sqrt{2f_0^2-1}}\) \no
&&-{1\over{f_0}}\sqrt{\frac{(f_0^2-1)^3}{2f_0^2-1}}
\Pi\(\sqrt{\frac{f^2-f_0^2}{f^2-1}},{1\over{f_0^2}},\frac{f_0}{\sqrt{2f_0^2-1}}\)
\ee

$\tau$ and $y$ are proper time and space-time rapidity. The $y=Y$ limit
of (\ref{eq:scaling}) relates the parameter $f_0$ and the quark rapidity
$Y=arctanh(v)$. It is also very useful to write down the EOM of $f(y)$:
\be\label{eq:f_eom}
f'=\frac{\sqrt{V(V-E^2)}}{E}
\ee
with $V=f^4-f^2$, $f_0^4-f_0^2-E^2=0$.

We want the source term due to the scaling string. It is convenient to 
switch to a different parametrization:
\be
z=z\(t,x\)\; , x_{\perp}=0
\ee
where $x$ and $x_{\perp}$ represent longitudinal and transverse coordinates
respectively. 
The above parametrization leads directly to

\begin{widetext}
\be\label{eq:source}
&&\dlt S_{\mu\nu}=-\frac{1}{3}\frac{\ka^2 z}{2\pi\alpha'}\dlt\(z-\bar z\)\dlt\(x^2\)\dlt\(x^3\)
\frac{1}{\sqrt{1-z_t^2+z_x^2}}\cdot \no
&&\begin{pmatrix}
1+2z_t^2+z_x^2& 3z_tz_x& -3z_t& & \\
3z_tz_x& z_t^2+2z_x^2-1& -3z_x& & \\
-3z_t& -3z_x& 2+z_t^2-z_x^2& & \\
& & & 2(1-z_t^2+z_x^2)& \\
& & & & 2(1-z_t^2+z_x^2)
\end{pmatrix}
\ee
\end{widetext}
(In the matrices here and below we only show the
nonzero enties: the adopted order of coordinate
indices is $t,z,x^1,x^2,x^3$.
)
With the help of string EOM, (15) of \cite{KMT}, $h$ can be expressed in a
very compact form: $h=-{2\over3}\frac{\ka^2}{2\pi\alpha'}\dlt\(x^2\)\dlt\(x^3\)
\frac{z_t^2-z_x^2+2}{\sqrt{1-z_t^2+z_x^2}}{1\over2}(z^2-\bar z^2)\theta(z-\bar z)$
We also record the generalized source for later reference:

\be\label{eq:smn_tx}
&&s_{mn}={1\over3}\frac{-\ka^2}{2\pi\alpha'}\dlt(x^2)\dlt(x^3)
\biggl[\frac{\bar z}{\sqrt{1-z_t^2+z_x^2}}\dlt(z-\bar z)\times \no
&&\begin{pmatrix}
1+2z_t^2+z_x^2& 3z_tz_x& &\\
3z_tz_x& z_t^2+2z_x^2-1& &\\
& & 2(1-z_t^2+z_x^2)& \\
& & & 2(1-z_t^2+z_x^2)
\end{pmatrix} \no
&&+\begin{pmatrix}
2\del_t& \del_x& \del_{x_2}& \del_{x_3}\\
\del_x& & & \\
\del_{x_2}& & & \\
\del_{x_3}& & &
\end{pmatrix}\frac{3\bar z z_t}{\sqrt{1-z_t^2+z_x^2}}\theta(z-\bar z) \no
&&+\begin{pmatrix}
& \del_t& & \\
\del_t& 2\del_x& \del_{x_2}& \del_{x_3}\\
& \del_{x_2}& & \\
& \del_{x_3}& &
\end{pmatrix}\frac{3\bar z z_x}{\sqrt{1-z_t^2+z_x^2}}\theta(z-\bar z) \no
&&+\begin{pmatrix}
\del_t^2& \del_t\del_x& \del_t\del_{x_2}& \del_t\del_{x_3}\\
\del_t\del_x& \del_x^2& \del_x\del_{x_2}& \del_x\del_{x_3}\\
\del_t\del_{x_2}& \del_x\del_{x_2}& \del_{x_2}^2& \del_{x_2}\del_{x_3}\\
\del_t\del_{x_3}& \del_x\del_{x_3}& \del_{x_2}\del_{x_3}& \del_{x_3}^2
\end{pmatrix}\frac{z_t^2-z_x^2+2}{\sqrt{1-z_t^2+z_x^2}}\times \no
&&{1\over2}(z^2-\bar z^2)\theta(z-\bar z)+
\begin{pmatrix}
-1& & & \\
& 1& & \\
& & 1& \\
& & & 1
\end{pmatrix}\frac{z_t^2-z_x^2+2}{\sqrt{1-z_t^2+z_x^2}}\times \no
&&\theta(z-\bar z)
\biggr]
\ee

With the source now at hand, we  proceed to the calculation of
stress tensor. We use similar substitutions as before:
$\overrightarrow\del_x=-\overleftarrow\del_x=\overleftarrow\del_x'$. Performing
the derivative explicitly, we find the z-integral and $x_{\perp}$-integral
can be done easily.
We arrive at the following result:
\begin{widetext}
\be\label{eq:Qmn}
&&Q_{mn}={1\over3}\frac{-\ka^2}{2\pi\alpha'}\int dtdx
\biggl[\frac{\bar z^2}{\sqrt{1-z_t^2+z_x^2}}P
\begin{pmatrix}
1+2z_t^2+z_x^2& 3z_tz_x& & \\
3z_tz_x& z_t^2+2z_x^2-1& & \\
& & 2(1-z_t^2+z_x^2)& \\
& & & 2(1-z_t^2+z_x^2)
\end{pmatrix}+ \no
&&\begin{pmatrix}
2(t'-t)& -(x'-x)& -x_2'& -x_3'\\
-(x'-x)& & & \\
-x_2'& & & \\
-x_3'& & &
\end{pmatrix}\frac{3\bar z z_t}{\sqrt{1-z_t^2+z_x^2}}P+
\begin{pmatrix}
& t'-t& & \\
t'-t& -2(x'-x)& -x_2'& -x_3'\\
& -x_2'& & \\
& -x_3'& &
\end{pmatrix}\frac{3\bar z z_x}{\sqrt{1-z_t^2+z_x^2}}P+ \no
&&\begin{pmatrix}
(t'-t)^2& -(t'-t)(x'-x)& -(t'-t)x_2'& -(t'-t)x_3'\\
-(t'-t)(x'-x)& (x'-x)^2& (x'-x)x_2'& (x'-x)x_3'\\
-(t'-t)x_2'& (x'-x)x_2'& x_2'^2& x_2'x_3'\\
-(t'-t)x_3'& (x'-x)x_3'& x_2'x_3'& x_3'^2
\end{pmatrix}\
\frac{z_t^2-z_x^2+2}{\sqrt{1-z_t^2+z_x^2}}P
\biggr]
\ee
with 
\be
&&P=\frac{12i}{(2\pi)^2}\biggl[
\frac{1}{((t'-t)^2-(x'-x)^2-x_{\perp}'^2-\bar z^2+i\epsilon)^4}
-\frac{1}{((t'-t)^2-(x'-x)^2-x_{\perp}'^2-\bar z^2-i\epsilon)^4}\biggr]  \no
&&\equiv\frac{12i}{(2\pi)^2}
\frac{\pm 1}{((t'-t)^2-(x'-x)^2-x_{\perp}'^2-\bar z^2\pm i\epsilon)^4}
\theta(t'-t)
\ee
\end{widetext}
which is just the integrated propagator.
The four matrices in the expression above we will refer to later
as I,II,III,IV, respectively.

Here we  replaced the theta function of $P_R$ by $\theta(t'-t)$. It is
justified since
the $\pm i\epsilon$ prescription  encodes 
derivatives of 
the delta function, and the theta function picks up only one pole
of the propagator.

In order to plugin the scaling solution
for the string, we return to
$\tau,y$ coordinates:

\be
&&z_t=\frac{chy}{f}+\frac{shyf'}{f^2} \no
&&z_x=-\frac{shy}{f}-\frac{chyf'}{f^2} \no
&&\int dtdx=\int\tau d\tau dy \nonumber
\ee

The source together with the integration measure has one of the
following simple $\tau$-dependence: $\tau,\tau^2,\tau^3$.  The
propagator now is

\be
&&P=\frac{12i}{(2\pi)^2}
\frac{\pm \theta(\tau'-\tau)}{((1-\frac{1}{f^2})\tau^2+\tau'^2-2\tau\tau'ch(y-y')-x_{\perp}'^2\pm i\epsilon)^4} \no
&&=\frac{12i}{(2\pi)^2}
\frac{\pm 1}{((1-\frac{1}{f^2})(\tau-\tau_+)(\tau-\tau_-)\pm i\epsilon)^4}\theta(\tau'-\tau)
\ee
with \be \tau_\pm=\frac{\tau'ch(y'-y)\pm\sqrt{\tau'^2ch^2(y'-y)-
(\tau'^2-x_{\perp}'^2)(1-\frac{1}{f^2})}}{1-\frac{1}{f^2}} \ee
This propagator as a function of $\tau$ 
contains two fourth order poles, so  
the $\tau$-integral is calculated by the residue theorem. Note that
the 
theta function picks up only one pole at $\tau=\tau_-$.
Since our $\tau$-integral extends from zero to infinity, we must have a
 positive $\tau_-$ in order to have a nonvanishing result, which requires
$\tau'^2-x_{\perp}'^2=t'^2-r'^2>0$, precisely the condition that the observer must
stay inside the lightcone.
Since
 the quark and the antiquark are emerging from the space-time origin,
  the stress tensor is indeed expected
to vanish outside the lightcone.

Completed the $\tau$ integral and replacing
 ${1\over3}\frac{-\ka^2}{2\pi\alpha'}$
by $\frac{-\sqrt{\lam}}{3\pi}$, we 
 convert $Q_{mn}$ to the boundary stress tensor
$T_{mn}$ (compare \cite{Lin:2007pv}). We thus get our final result 

\begin{widetext}
\be\label{eq:fint}
&&T_{mn}=\frac{-\sqrt{\lam}}{3\pi}\int_{-Y}^{Y}dy
\biggl[
\begin{pmatrix}
1+2z_t^2+z_x^2& 3z_tz_x& & \\
3z_tz_x& z_t^2+2z_x^2-1& & \\
& & 2(1-z_t^2+z_x^2)& \\
& & & 2(1-z_t^2+z_x^2)
\end{pmatrix}\frac{1}{f^2\sqrt{1-z_t^2+z_x^2}}A \no
&&+\begin{pmatrix}
2t'& -x'& -x_2'& -x_3'\\
-x'& & & \\
-x_2'& & & \\
-x_3'& & & 
\end{pmatrix}\frac{3z_t}{f\sqrt{1-z_t^2+z_x^2}}B+
\begin{pmatrix}
-2chy& shy& 0& 0\\
shy& & & \\
0& & & \\
0& & &
\end{pmatrix}\frac{3z_t}{f\sqrt{1-z_t^2+z_x^2}}A \no
&&+\begin{pmatrix}
& t'& & \\
t'& -2x'& -x_2'& -x_3'\\
& -x_2'& & \\
& -x_3'& &
\end{pmatrix}\frac{3z_x}{f\sqrt{1-z_t^2+z_x^2}}B+
\begin{pmatrix}
& -chy& & \\
-chy& 2shy& 0& 0\\
& 0& & \\
& 0& &
\end{pmatrix}\frac{3z_x}{f\sqrt{1-z_t^2+z_x^2}}A \no
&&+\begin{pmatrix}
t'^2& -t'x'& -t'x_2'& -t'x_3'\\
-t'x'& x'^2& x'x_2'& x'x_3'\\
-t'x_2'& x'x_2'& x_2'^2& x_2'x_3'\\
-t'x_3'& x'x_3'& x_2'x_3'& x_3'^2
\end{pmatrix}\frac{z_t^2-z_x^2+2}{\sqrt{1-z_t^2+z_x^2}}C+
\begin{pmatrix}
-2t'chy& t'shy+x'chy& x_2'chy& x_3'chy\\
t'shy+x'chy& -2x'shy& -x_2'shy& -x_3'shy\\
x_2'chy& -x_2'shy& & \\
x_3'chy& -x_3'shy& &
\end{pmatrix}\times \no
&&\frac{z_t^2-z_x^2+2}{\sqrt{1-z_t^2+z_x^2}}B
+\begin{pmatrix}
chy^2& -chyshy& & \\
-chyshy& shy^2& & \\
& & & \\
& & &
\end{pmatrix}\frac{z_t^2-z_x^2+2}{\sqrt{1-z_t^2+z_x^2}}A
\biggr]
\\
&&A={1\over{(1-\frac{1}{f^2})^4}}\frac{\tp^3+9\tp^2\tm+9\tp\tm^2+\tm^3}{(\tp-\tm)^7}\no
&&B={4\over{(1-\frac{1}{f^2})^4}}\frac{\tp^2+3\tp\tm+\tm^2}{(\tp-\tm)^7}\no
&&C={10\over{(1-\frac{1}{f^2})^4}}\frac{\tp+\tm}{(\tp-\tm)^7}\no
\ee
\end{widetext}

\subsection{A field near one charge}

We first consider the stress tensor close to the quark. Note the quark moves
with velocity v, the observer should also move in order to stay close to
the quark. Thus it is convenient to switch to rest coordinate of the 
moving quark, yet we still stay in the rest frame of the system. The rest
coordinates of the quark, indicated by a tilde,
 relates the original coordinates in the following way:

\be
&&\tlt=chYt'-shYx'\no
&&\tlx=chYx'-shYt'\no
&&\tlx_2=x_2'\no
&&\tlx_3=x_3'
\ee 

We expect to obtain the field by the quark only, provided we are sufficient 
close to the quark, which corresponds to the limit
 $\tlt\gg\tlr\; ,\tlr\equiv\sqrt{\tlx^2+\tlx_2^2+\tlx_3^2}\rightarrow 0$. Comparing the string profile for the
stretching dipole with that for a single quark, we claim the leading order
field near the quark receives contribution from the quark end of
the string, which corresponds to the integration of $y$ near
$y=Y$ in (\ref{eq:fint}). If we instead do the integral in $f$ 
with $dy=\frac{df}{f'}$, we may only focus on large $f$ integration.

Note $f'\sim {f^4\over E}\;y-Y\sim{1\over f^3}$. To the leading order, 
we can simplfy
replace $y$ by $Y$, which leads to the following relations:

\be
&&\tp-\tm=\frac{2\sqrt{\tlt^2-(1-\frac{1}{f^2})(\tlt^2-\tlr^2)}}{1-\frac{1}{f^2}}\no
&&\tp^3+9\tp^2\tm+9\tp\tm^2+\tm^3=\frac{8\tlt^3}{(1-\frac{1}{f^2})^3}+ \no
&&\frac{12(\tlt^2-\tlr^2)\tlt}{(1-\frac{1}{f^2})^2}\no
&&\tp^2+3\tp\tm+\tm^2=\frac{4\tlt^2}{(1-\frac{1}{f^2})^2}+\frac{\tlt^2-\tlr^2}{1-\frac{1}{f^2}}\no
&&\tp+\tm=\frac{2\tlt}{1-\frac{1}{f^2}}
\ee

Performing the $f$ integral near $\infty$, and take the limit
$\tlt\gg\tlr\; ,\tlr\rightarrow 0$, which is essentially a small $\tlr$
expansion, we find the leading order field given by source I and IV
diverges as $O({1\over \tlr^4})$, while source II and III only yield
subleading contribution $O({1\over\tlr^2})$. We display the LO
field near the quark as follows:

\be\label{eq:LO}
&&T_{mn}=\frac{2\sqrt{\lam}}{\pi^2}
\biggl[
\begin{pmatrix}
1+2\gm^2\bt^2& -2\gm^2\bt& & \\
-2\gm^2\bt& -1+2\gm^2& & \\
& & 1& \\
& & & 1
\end{pmatrix}\frac{1}{24\tlr^4}- \no
&&\begin{pmatrix}
\gm^2\bt^2\tlx^2& -\gm^2\bt\tlx^2& -\gm\bt\tlx\tlx_2& -\gm\bt\tlx\tlx_3\\
-\gm^2\bt\tlx^2& \gm^2\tlx^2& \gm\tlx\tlx_2& \gm\tlx\tlx_3\\
-\gm\bt\tlx\tlx_2& \gm\tlx\tlx_2& \tlx_2^2& \tlx_2\tlx_2\\
-\gm\bt\tlx\tlx_3& \gm\tlx\tlx_3& \tlx_2\tlx_3& \tlx_3^2
\end{pmatrix}\frac{1}{12\tlr^6}
\biggr] \no
\ee

with $\gm=chY\; ,\gm\bt=shY$.

This does not look very nice at first glance, actually it is just the
stress tensor of a static quark boosted to a frame moving with velocity
$-v$. It is clearly traceless and divergence-free. We confirm that
the LO field near the quark contains contribution from the quark only.

Next we would like to extend the result to NLO to include the effect
of the antiquark. Note there are two possible corrections relevant
for NLO: correction to the source $\Delta y=y-Y={\Delta f\over f'}=
{E\over {3f^3}}$, thus $chy=chY+shY\Delta y$, $shy=shY+chY\Delta y$.
The other is correction to the propagator: 
\be
&&P=\frac{12i}{(2\pi)^2}
\frac{\pm \theta(\tau'-\tau)}{((1-\frac{1}{f^2})\tau^2+\tau'^2-2\tau\tau'ch(y-y')-x_{\perp}'^2\pm i\epsilon)^4} \no
&&=\frac{12i}{(2\pi)^2}\biggl[
\frac{\pm 1}{((1-\frac{1}{f^2})\tau^2+\tau'^2-2\tau\tau'ch(y'-Y)-x_{\perp}'^2\pm i\epsilon)^4} \no
&&+\frac{\pm 1}{((1-\frac{1}{f^2})\tau^2+\tau'^2-2\tau\tau'ch(y'-Y)-x_{\perp}'^2\pm i\epsilon)^5}\times \no
&&(-8\tau\tau'sh(y'-Y)\Delta y)+\cdots\biggr]\theta(\tau'-\tau) \no
&&=\frac{12i}{(2\pi)^2}\big[
\frac{\pm 1}{((1-\frac{1}{f^2})\tau^2+\tau'^2-2\tau\tau'ch(y'-Y)-x_{\perp}'^2\pm i\epsilon)^4} \no
&&+\frac{\pm 1}{((1-\frac{1}{f^2})\tau^2+\tau'^2-2\tau\tau'ch(y'-Y)-x_{\perp}'^2\pm i\epsilon)^5}(\frac{8E\tau\tlx}{3f^3}) \no
&&+\cdots
\big]\theta(\tau'-\tau) \no
&&=\frac{12i}{(2\pi)^2}[P_4+P_5+\cdots]\theta(\tau'-\tau)
\ee

with
\be
&&P_4=
\frac{\pm 1}{((1-\frac{1}{f^2})\tau^2+\tau'^2-2\tau\tau'ch(y'-Y)-x_{\perp}'^2\pm i\epsilon)^4} \no
&&P_5=
\frac{\pm 1}{((1-\frac{1}{f^2})\tau^2+\tau'^2-2\tau\tau'ch(y'-Y)-x_{\perp}'^2\pm i\epsilon)^5}(\frac{8E\tau\tlx}{3f^3}) \nonumber
\ee

After relatively lengthy calculation and comparison we find
that the NLO correction is composed of three pieces: the first one is
source II and III convoluted with $P_4$, the second
 correspond to source IV
and $P_4$, and the third one comes from
the convolusion of source I and $P_5$. Collecting all of them,
we find the following result:

\be\label{eq:NLO}
&&T_{mn}=\frac{2\sqrt{\lam E}}{\pi^2}
\biggl[
-\begin{pmatrix}
5+13\gm^2\bt^2& -13\gm^2\bt& & \\
-13\gm^2\bt& -5+13\gm^2& & \\
& & 8& \\
& & & 8
\end{pmatrix}\frac{\tlx}{144\tlr^3\tlt^2} \no
&&+\begin{pmatrix}
\gm^2\bt^2\tlx^2& -\gm^2\bt\tlx^2& -\gm\bt\tlx\tlx_2& -\gm\bt\tlx\tlx_3\\
-\gm^2\bt\tlx^2& \gm^2\tlx^2& \gm\tlx\tlx_2& \gm\tlx\tlx_3\\
-\gm\bt\tlx\tlx_2& \gm\tlx\tlx_2& \tlx_2^2& \tlx_2\tlx_3\\
-\gm\bt\tlx\tlx_3& \gm\tlx\tlx_3& \tlx_2\tlx_3& \tlx_3^2
\end{pmatrix}\frac{\tlx}{48\tlr^5\tlt^2} \no
&&+\begin{pmatrix}
2\gm^2\bt^2\tlx& -2\gm^2\bt\tlx& -\gm\bt\tlx_2& -\gm\bt\tlx_3\\
-2\gm^2\bt\tlx& 2\gm^2\tlx& \gm\tlx_2& \gm\tlx_3\\
-\gm\bt\tlx_2& \gm\tlx_2& & \\
-\gm\bt\tlx_3& \gm\tlx_3& &
\end{pmatrix}\frac{1}{18\tlr^3\tlt^2}
\biggr]
\ee
We had found with satisfaction that this result is
indeed traceless and divergence-free (to order $O(\frac{1}{\tlr^3})$).

After this result is boosted to a frame moving with velocity $+v$,
it reproduces the NLO field near the quark of a static dipole with
the identification $\frac{E}{\tlt^2}\rightarrow\frac{1}{z_m^2}\approx
\frac{1.2^2}{L^2}$($L$ is the quark-antiquark seperation).
It confirms that close to one charge it is not important to this
accuracy what the other charge is doing.
 In fact in the quasi-static limit $v\rightarrow 0$, 
$\frac{E}{\tlt^2}\approx\frac{1.2^2}{4v^2\tlt^2}=\frac{1.2^2}{\tilde{L}^2}$,
 where $\tilde{L}$ is the quark-antiquark seperation at time $\tlt$.

\subsection{The slow-moving limit}

Since the stretching string solution depends
on only one parameter $v$, the ends velocity, which is bounded
from above by the critical velocity.  One interesting limit
in which calculations can be pushed a step further is
$v\rightarrow 0$, which correspond to slow motion.
 In practice, it corresponds
to
expantion of  the stress tensor in inverse powers of $f_0$.

We start with considering the large $f_0$ limit of (\ref{eq:fint}). 
Define $\eta=\frac{f}{f_0}\;\eta\geqslant 1$ such that the range of
$\eta$ is independent from $f_0$. The large $f_0$ expansion of $y$ is
a little complicated:

\be
&&y={1\over2}\sqrt{f_0^4-f_0^2}\biggl[\frac{2{\it F}(\frac{f^2-f_0^2}{f^2-1},\frac{f_0^2}{2f_0^2-1})}{\sqrt{2f_0^2-1}} \no
&&-\frac{2(f_0^2-1){\it \Pi}(\frac{f^2-f_0^2}{f^2-1},\frac{1}{f_0^2},\frac{f_0^2}{2f_0^2-1})}{\sqrt{2f_0^2-1}f_0^2}\biggr] \no
&&=\frac{-{\it F}(\frac{\sqrt{\eta^2-1}}{\eta},\frac{\sqrt{2}}{2})
+2{\it E}(\frac{\sqrt{\eta^2-1}}{\eta},\frac{\sqrt{2}}{2})}{\sqrt{2}f_0}
+O(\frac{1}{f_0^3}) \no
&&=\frac{G(\eta)}{f_0}+O(\frac{1}{f_0^3})
\ee
where $G(\eta)=\frac{-{\it F}(\frac{\sqrt{\eta^2-1}}{\eta},\frac{\sqrt{2}}{2})
+2{\it E}(\frac{\sqrt{\eta^2-1}}{\eta},\frac{\sqrt{2}}{2})}{\sqrt{2}}$

With the asymptotic expansion of $y$ and $f_0$, we are ready to proceed to
the stress tensor. It seems at first glance the leading order is of 
$O({1\over {f_0}})$, given by source IV. Actually the prefactor, which is 
an integral of $\eta$ vanishes. The order $O({1\over {f_0^2}})$ does not
contribute either due to the symmetry $y\leftrightarrow -y$. Finally we
have to extend the calculation to order $O({1\over {f_0^3}})$. Expand all the
relevant quantity in $f_0$, and keep the order $O({1\over {f_0^3}})$ in the
result of stress tensor. It is a quite lengthy but straight forward 
calculation. The result is displayed as follows(we have omitted the prime
in boundary coordinates):

\begin{widetext}
\be\label{eq:quasistatic}
&&T_{tt}=\frac{2\sqrt{\lam}}{f_0^3\pi^2}\frac{2t}{r^9}\big[(10a+5e_1+5e_2)r^2t^2+45e_1x^2r^2-35e_1t^2x^2-(9e_2+2f+6a-24c+7e_1)r^4\big] \no
&&T_{tx}=\frac{2\sqrt{\lam}}{f_0^3\pi^2}\frac{2x}{r^9}\big[(45e_1+15e_2-90d-30c)r^2t^2+15e_1x^2r^2-105e_1t^2x^2-(3e_2+7e_1+2f-6c-18d)r^4\big] \no
&&T_{tx_i}=\frac{2\sqrt{\lam}}{f_0^3\pi^2}\frac{2x_i}{r^9}\big[(15e_2+15e_1-30c)r^2t^2-105e_1t^2x^2+15e_1x^2r^2\big] \no
&&T_{xx}=\frac{2\sqrt{\lam}}{f_0^3\pi^2}\frac{2t}{r^{11}}\big[(20a-30b+10e_1-60d)r^4t^2+(-12a+18b-6e_1+36d)r^6+(420d-175e_1-35e_2)r^2t^2x^2 \no
&&+(-180d+15e_2+65e_1+10f)x^2r^4+315e_1x^4t^2-105e_1x^4r^2\big] \no
&&T_{xx_i}=\frac{2\sqrt{\lam}}{f_0^3\pi^2}\frac{10txx_i}{r^{11}}\big[(42d-21e_1-7e_2)t^2r^2-21e_1x^2r^2+63e_1t^2x^2\big] \no
&&T_{x_ix_j}=\frac{2\sqrt{\lam}}{f_0^3\pi^2}\frac{8t}{r^7}(5at^2-3ar^2)\dlt_{ij}-\frac{2\sqrt{\lam}}{f_0^3\pi^2}\frac{10tx_ix_j}{r^{11}}
\big[(7e_1+7e_2)r^2t^2-(3e_2+e_1+2f)r^4+21e_1x^2r^2 \no
&&-63e_1x^2t^2\big]
\ee
with
\be
&&r=\sqrt{x^2+x_2^2+x_3^2} \no
&&a=\int_1^\infty\frac{1}{\eta^2\sqrt{\eta^4-1}}d\eta=0.5991 \no
&&b=\int_1^\infty\frac{1}{\eta^6\sqrt{\eta^4-1}}d\eta=0.3594 \no
&&c=\int_1^\infty({1\over\eta}+G(\eta)\sqrt{\eta^4-1})
\frac{1}{\eta^5\sqrt{\eta^4-1}}d\eta=0.4493 \no
&&d=\int_1^\infty\frac{G(\eta)}{\eta^5}d\eta=0.0899 \no
&&e_1=\int_1^\infty\frac{3-\eta^4}{\eta^4\sqrt{\eta^4-1}}G(\eta)^2d\eta=-0.1797 \no
&&e_2=\int_1^\infty\frac{3-\eta^4}{\eta^6\sqrt{\eta^4-1}}d\eta=0.4793 \no
&&f=\int_1^\infty\frac{-5\eta^4+6\eta^2+9}{2\sqrt{\eta^4-1}(\eta^2+1)\eta^6}d\eta=0.7189 \no
\ee
\end{widetext}

Several comments about the result are in order:
(i)the result applies for arbitrary point on the boundary, 
i.e.general $t,x,x_2,x_3$, provided the point lies inside the lightcone. The
discontinuity of the stress tensor on the lightcone 
is a consequence of the discontinuity in
source at $t=0$
(ii)trace and divergence of the stress tensor vanish for any points away 
from the trajectory of the dipole ends, which is of course implicitly
assumed in our calculation.
(iii)If we consider the limit $t\gg r$, which amounts to keeping only the
highest power of $t$. Recalling the quasi-static 
limit: $\frac{t^3}{f_0^3}\sim(\frac{vt}{0.6})^3=(\frac{L}{1.2})^3$, where
$L$ is the dipole size at time $t$. While for the case of static
dipole: $z_m^3=(\frac{L}{1.2})^3$, we find the stretching dipole result
agrees with static dipole in the double limits: $v\rightarrow0$,$t\gg r$.
The numerical factors matches as well.
(vi)the agreement of quasi-static result and NLO near field with those of
static dipole by the simple identification: $L=2vt$ seems to suggest that
the vacuum-quark(antiquark) interaction is instantaneous.

\begin{figure}
\includegraphics[width=0.3\textwidth]{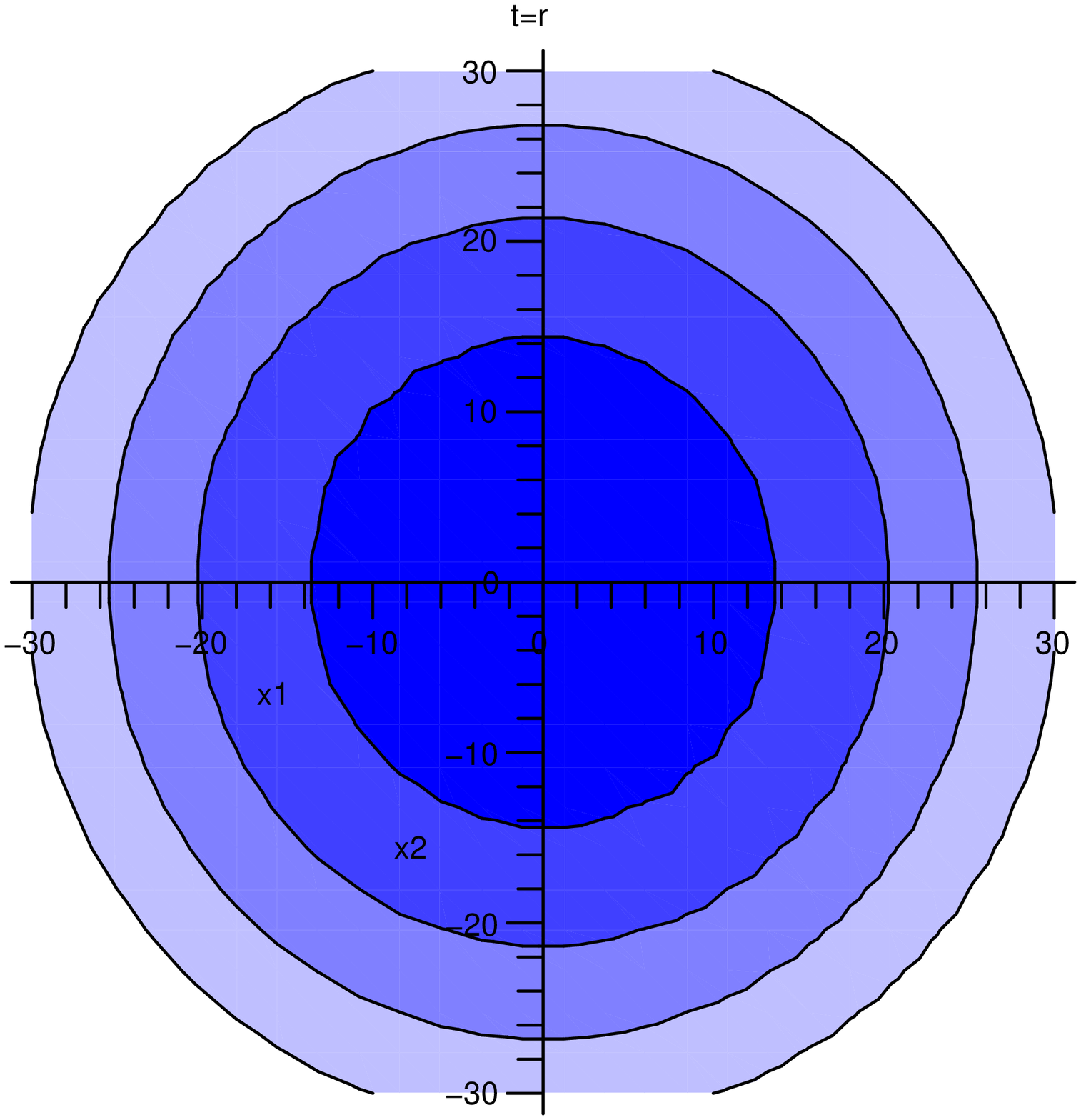}
\includegraphics[width=0.3\textwidth]{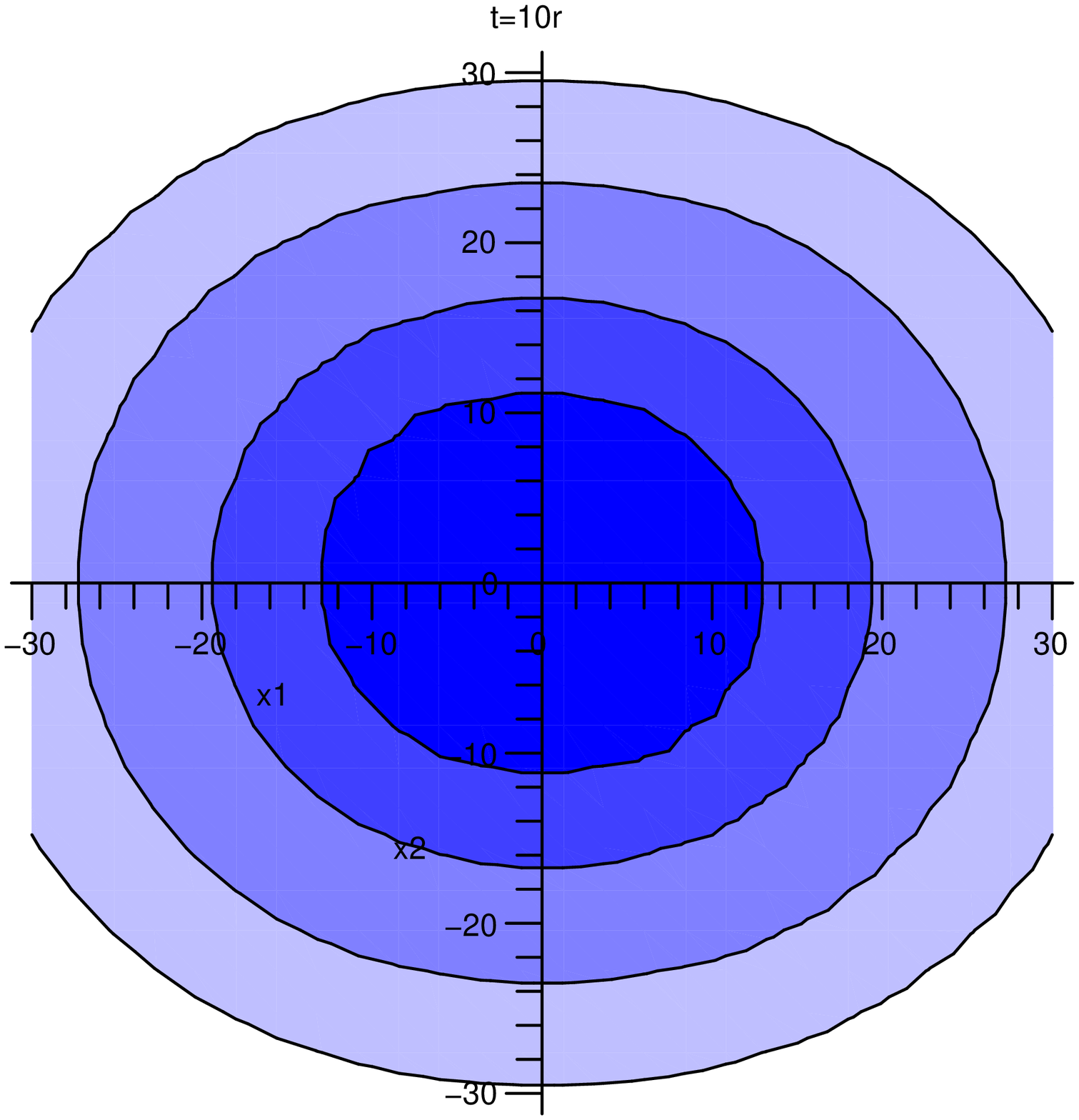}
\includegraphics[width=0.3\textwidth]{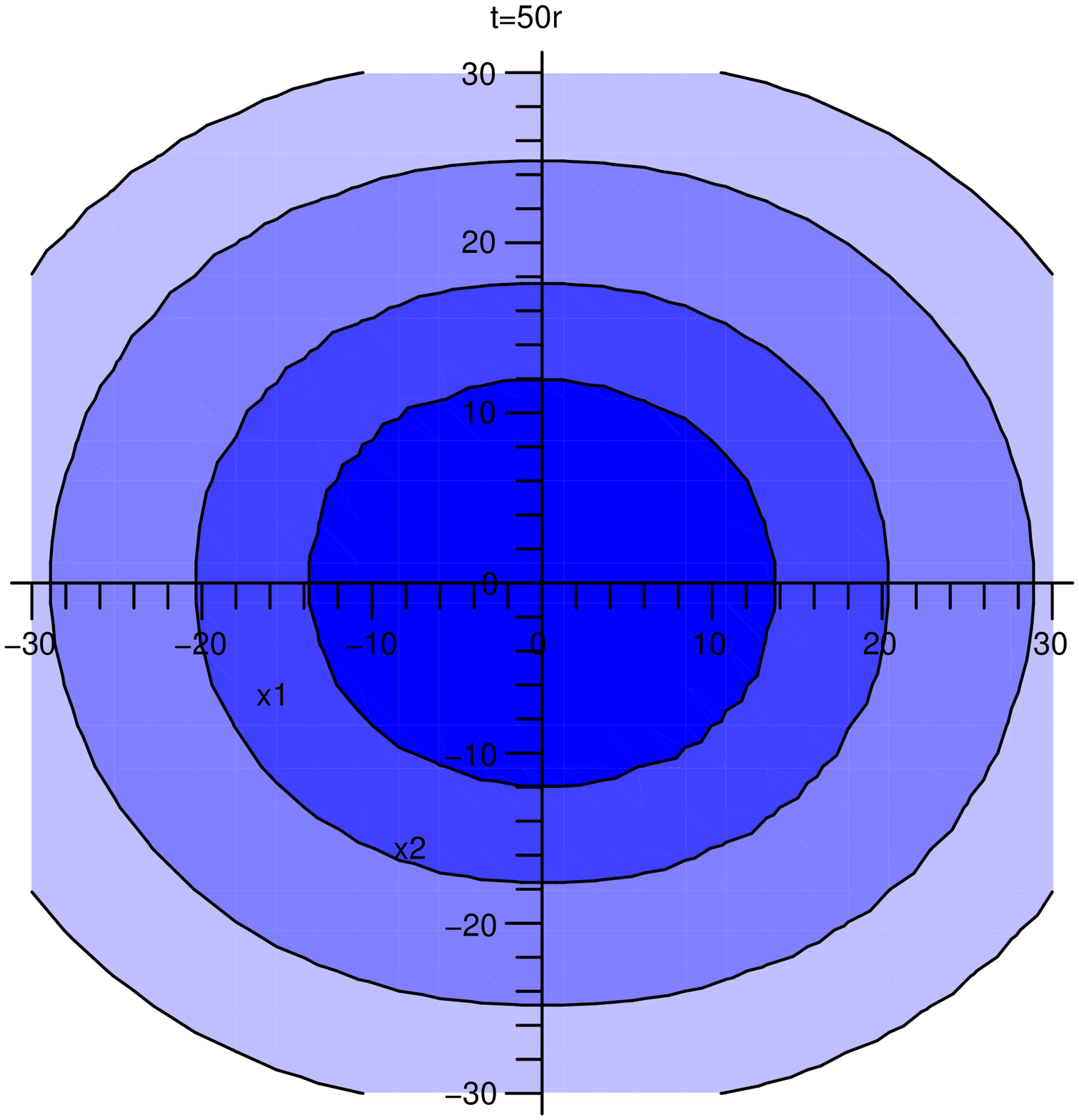}
\caption{\label{fig:e}(color online)
The contours of energy density $T^{00}$,
in unit of $\frac{2\sqrt{\lam}}{f_0^3\pi^2}$,
in $x_1-x_2$ plane at different time. The three plots are made for $t=r$,
$t=10r$ and $t=50r$ from top to bottom.
Note the quark/antiquark is at $x_1=\pm vt$. In the slow moving limit,
they are nearly at the origin.
The magnitude of $T^{00}$ is represented
by the color, with darker color corresponding to greater magnitude.
As time increases, the shape of the
contours gets elongated along $x_1$ axis}
\end{figure}

\begin{figure}
\includegraphics[width=0.3\textwidth]{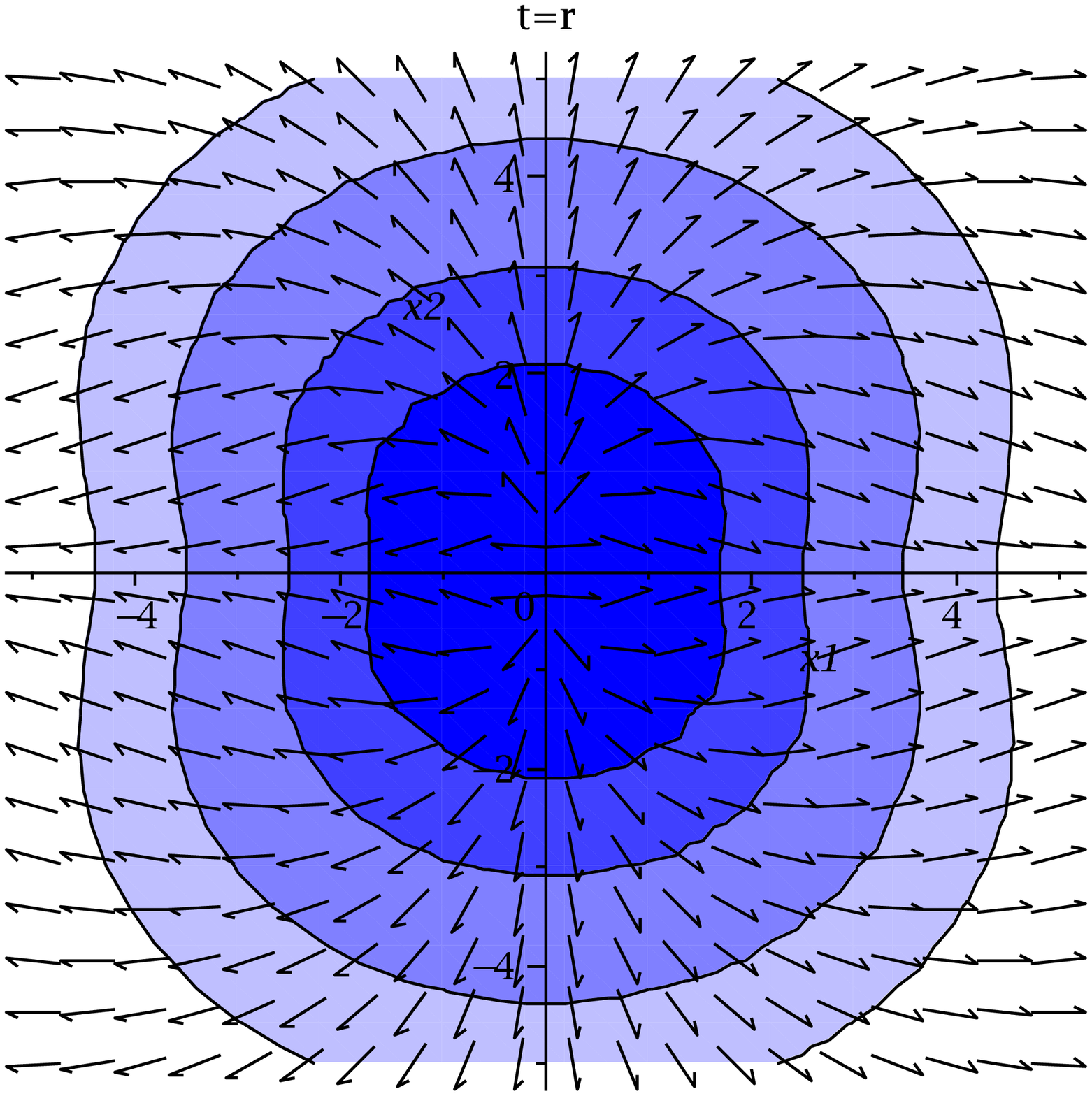}
\includegraphics[width=0.3\textwidth]{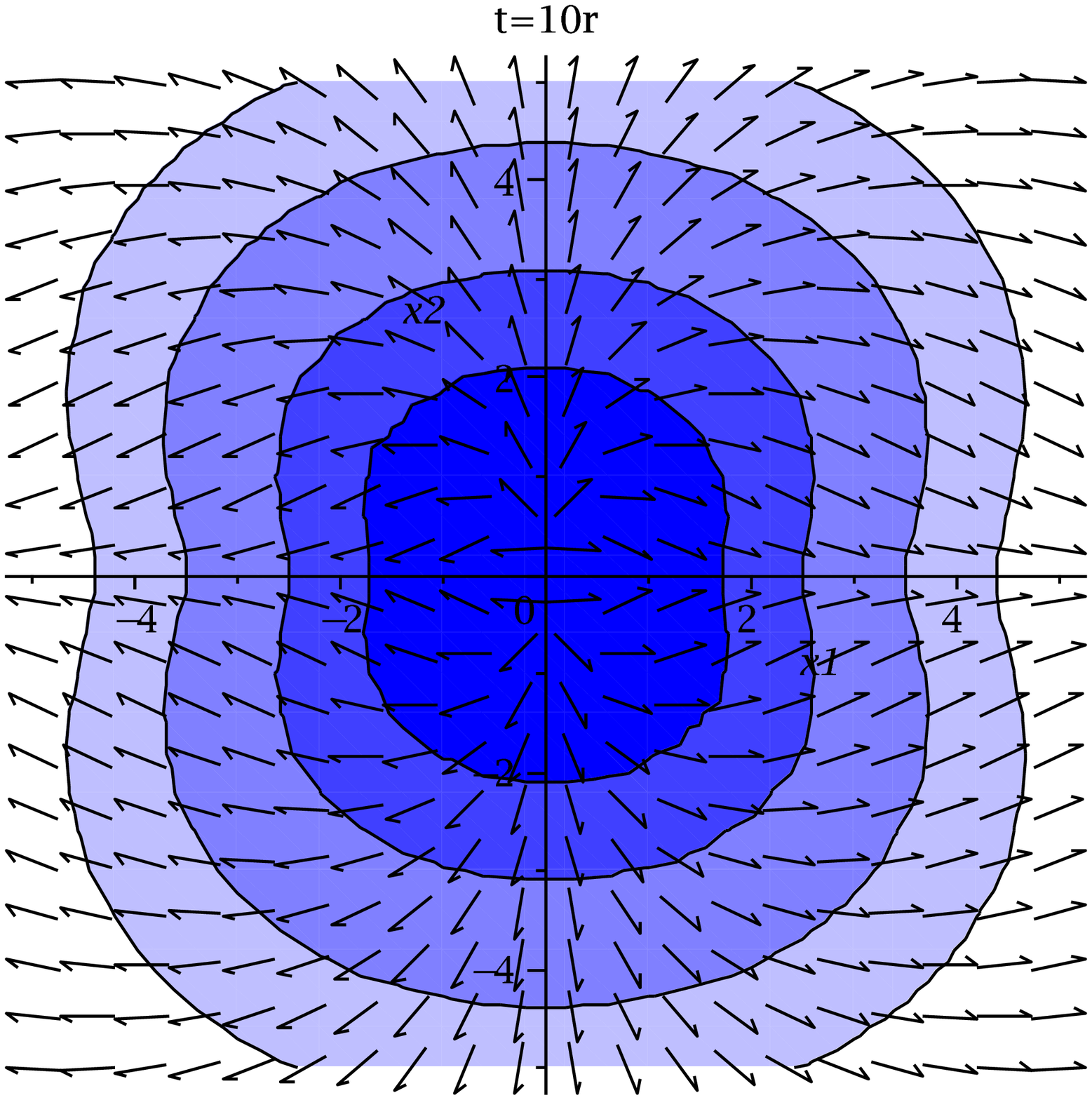}
\includegraphics[width=0.3\textwidth]{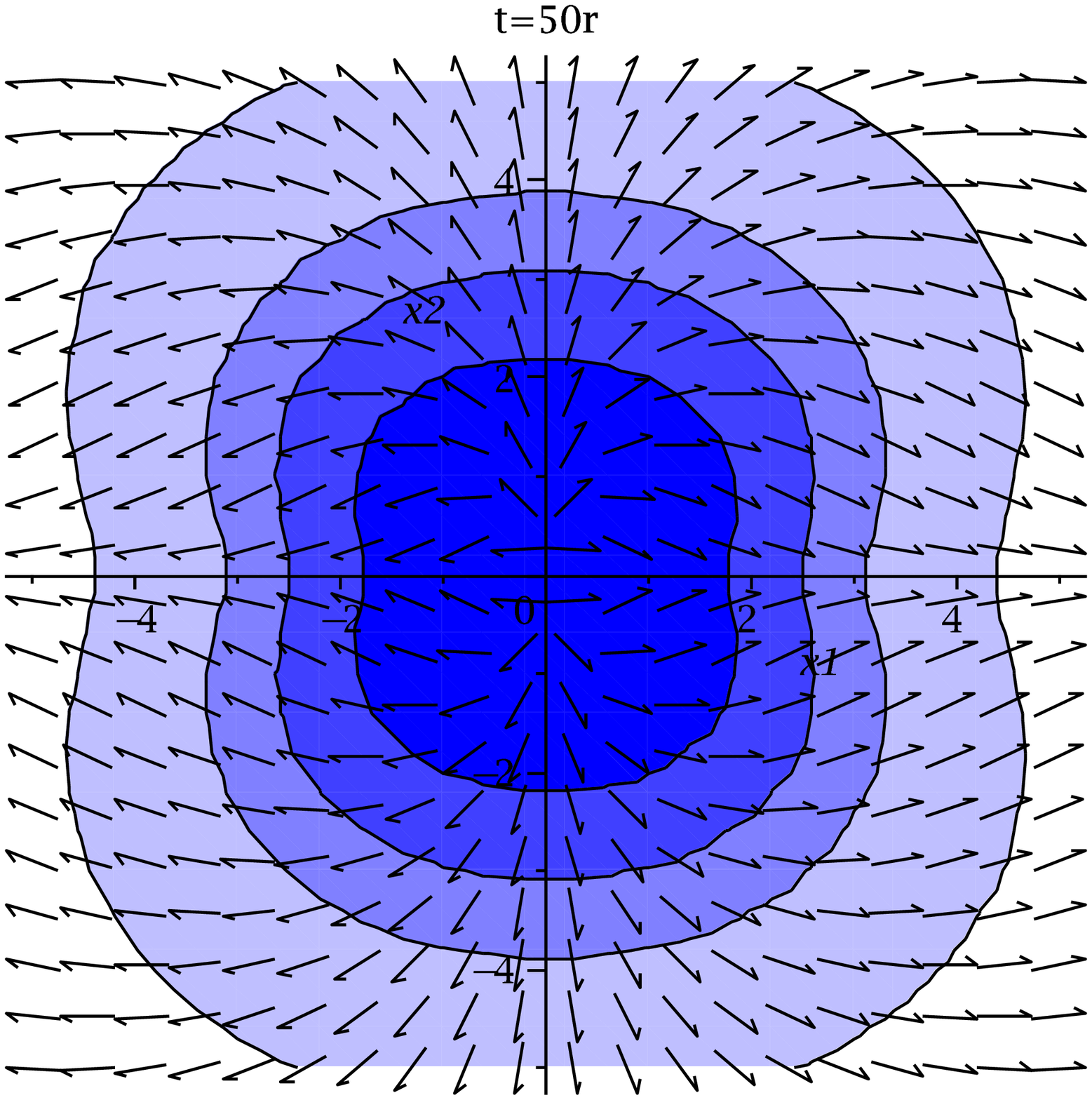}
\caption{\label{fig:p}(color online)The contours of momentum density
$T^{0i}$, in unit of $\frac{2\sqrt{\lam}}{f_0^3\pi^2}$,
in $x_1-x_2$ plane at different time. The three plots are made for $t=r$,
$t=10r$ and $t=50r$ from top to bottom.
Note the quark/antiquark is at $x_1=\pm vt$. In the slow moving limit,
they are nearly at the origin.
The magnitude is represented by color, with darker color corresponding to
 greater magnitude. 
The direction of the momentum density is indicated by normalized arrows}\end{figure}

We plot the energy density $T^{00}$ and
the momentum density (energy flow) $T^{0i}$ as a function
of the spatial coordinates at three different times
 in Fig.\ref{fig:e} and 
Fig.\ref{fig:p}, respectively. We observe that 
although the shape of energy distribution becomes
more alongated with time, reflecting changing shape of the string,
 the shape of the
 momentum flows seems to stay the same, with
 an interesting ``eight'' 
shape or forward-backward depletion. Small arrows 
display the direction of the energy flow.  Although they 
overall show outgoing explosion away from 
the origin (the collision point), one can also see some
``cumulative'' flow with jets converging along the collision axes.

\section{The energy density of matter in comoving frame and freezeout}

It is illuminating to ask if the stress tensor we obtained 
can or cannot
be described by some  hydrodynamical flow.
The latter is widely used in describing heavy ion collisions
\cite{Teaney:2001av,Kolb:2003dz}. More precisely, the question
is if  our stress tensor 
(\ref{eq:quasistatic})
is  that of a flowing
 ideal liquid \be T_{\mu\nu}=(\epsilon+p)u_{\mu}u_{\nu}
+pg_{\mu\nu} \ee where $u_{\mu}$ is the 4-velocity of the liquid and
 $\epsilon$
and $p$ are the energy density and the pressure. (Tracelessness
would of course demand that $\epsilon=3p$)
It is not difficult to show that this is $not$ the case:
the structure of our answer is richer than this simple form.

Nevertheless, it is still possible
to define a ``comoving frame'' of matter at any point,
in which the (boosted) momentum density $T_{0i}'$ vanishes.
 The (boosted) local value $T_{00}'$ component 
is the energy density in such a comoving frame,
 which we denote by $\epsilon$ like for a liquid.
 We use the contour of $\epsilon$ to define
the freezeout surface.

However the (boosted) local values of other spatial 
componets are in general unrelated and can be viewed
as ``anistropic pressure'' of non-equilibrium matter.
Needless to say, it remains unknown
 what combinations of fundamental fields
of the $\cal N$=4 theory -- the gauge one, the fermion or the
 scalars --
 participate in this flow of produced matter:
 to learn that one should do
``holography'' for
many more operators on the boundary.

Recall that the stress tensor in different
frames are related by $T_{\alpha\beta}'=S^{\mu}_{\alpha}S^{\nu}_{\beta}T_{\mu\nu}$ where
$S^{\mu}_{\alpha}=\frac{\del{x^{\mu}}}{\del{x^{\alpha}}}$ are 
Lorentz boost matrix.
The primed quantities correspond to new  frame.
Therefore the aim is to find
such boost which kills all the $(0,i)$ components.

 In practice, it is achieved numerically by the
following recipe:

We pick up any point inside the lightcone, calculate the eigenvalue and 
associate eigenvectors of the corresponding stress tensor matrix. Out of 
the four eigenvalues, one is selected to be the local energy density based on 
its eigenvector(See Appendix B for a short explanation). Fig.\ref{fig:3d} is a
surface plot of $\epsilon$ profile in spatial coordinate. The
 plot is made for $t=1$. It shows a nearly spherical shape for contour with
large $r$ and an elongated shape for contour with small $r$. By virtue of  
conformality of the setting, this translates to the following:
At early time the local energy density contour is nearly spherical
and at late time it gets elongated along $x_1$ axis.

\begin{figure}
\includegraphics[width=0.5\textwidth]{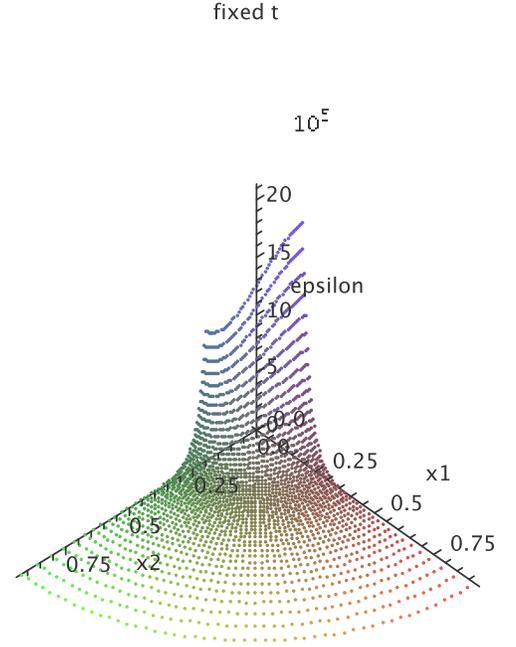}
\caption{\label{fig:3d}(color online)the profile of $\epsilon$,
in unit of $\frac{2\sqrt{\lam}}{f_0^3\pi^2}$, at $t=1$
with $r\approx 0.2-1$. The evolution of the shape of contour, i.e. the 
freezeout surface is contained in this
plot. The contours with large $r$(small $\epsilon$) are nearly spherical 
while the contours with small $r$(large $\epsilon$) are elongated along 
$x_1$ axis}
\end{figure}

\section{The stress tensor of multiple strings}

A simple extension of what is done above
is to consider many collidng quark-antiquark pairs 
uniformly distributed in the transverse plane. Every pair with the same
transverse coordinate is connected with a string. 
Let's assume quark and antiquark
only interact pairwise, which in the dual picture means the strings
do not interact with each other. As a result, the overall stress tensor
induced by the mutiple strings, in the linearized approximation simply
 amounts to integrating (\ref{eq:quasistatic}) over the transverse coordinates.
Note in order to preserve causality, the integral is done for 
$0<x_\perp^2=r^2-x^2<t^2-x^2$
We display the result as follows(we omit a factor of transverse string 
density, which does not alter space-time dependence of the stress tensor):


\be\label{eq:mstring}
&&T_{tt}=\frac{8\sqrt{\lam}}{f_0^3\pi}\big[-4e_1\frac{x^2}{t^4}+({4\over3}e_1
+{2\over3}f+2e_2-8c)\frac{1}{t^2} \no
&&+({20\over3}e_1-{2\over3}f+8c-3e_2-2a)
\frac{t}{x^3}+(e_2+2a-4e_1)\frac{t^3}{x^5}\big] \no
&&T_{tx}=\frac{8\sqrt{\lam}}{f_0^3\pi}\big[12e_1\frac{x^3}{t^5}+({-20\over3}e_1+4c+12d+{2\over3}f-2e_2)\frac{x}{t^3}+ \no
&&({2\over3}e_1+2c-{2\over3}f-e_2+6d)\frac{1}{x^2}+(-6c-6e_1-18d+3e_2)\frac{t^2}{x^4}\big] \no
&&T_{tx_i}=0 \no
&&T_{xx}=\frac{8\sqrt{\lam}}{f_0^3\pi}\big[-20e_1\frac{x^4}{t^6}+(12e_1+2e_2-2f-24d)\frac{x^2}{t^4} \no
&&+(6b-4a-4e_1+3e_2-24d+2f)\frac{t}{x^3}+(12e_1-6b+48d \no
&&+4a-5e_2)\frac{t^3}{x^5}\big] \no
&&T_{xx_i}=0 \no
&&T_{x_ix_j}=\frac{8\sqrt{\lam}}{f_0^3\pi}\big[10e_1\frac{x^4}{t^6}+(f-e_2-14e_1)\frac{x^2}{t^4} \no
&&+(e_2-{5\over3}f+{8\over3}e_1)\frac{1}{t^2}+(-4a+e_2+{2\over3}f-{8\over3}e_1)\frac{t}{x^3}+(4a \no
&&-e_2+4e_1)\frac{t^3}{x^5}\big] \no
\ee

\begin{figure}
\includegraphics[width=0.5\textwidth]{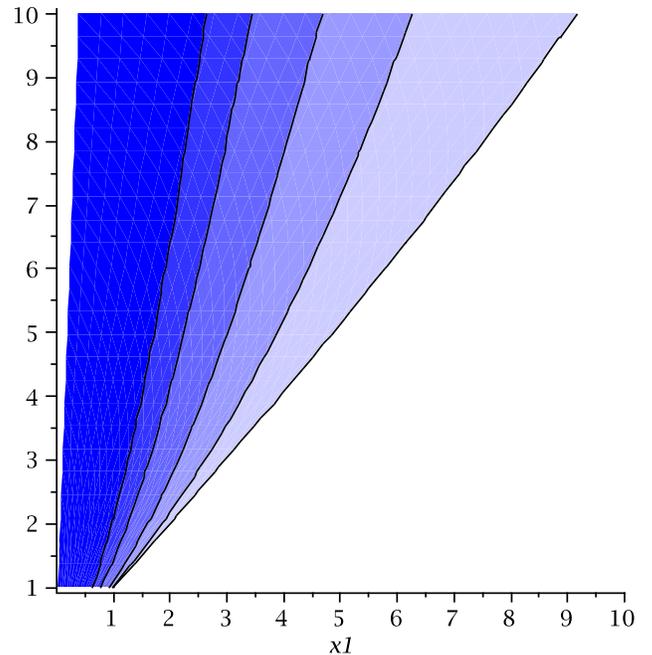}
\caption{\label{fig:int}(color online)The contours of energy density,
in unit of $\frac{8\sqrt{\lam}}{f_0^3\pi}$, as a function of $t$ and $x_1$.
The magnitude is represented by color, with darker color corresponding to
 greater magnitude. 
}
\end{figure}

We plot the energy density profiles in Fig.\ref{fig:int}.

\section{Summary and discussion}

The main objective of this paper was to calculate a ``hologram''
of the falling open string, which has ends attached to heavy quarks
moving with constant velocities $\pm v$. After an appropriate
  tool -- Green function for time-dependent linearized Einstein equation -- was
  constructed, and stress tensor of the string calculated, a convolution of the
  two gave us the stress tensor of  an ``explosion'' seen by an observer at
  the AdS boundary. Apart of analytical results in different limits, we have 
   given pictures of the time evolution of the energy
   density and the Poynting vector in Figs.\ref{fig:e}, \ref{fig:p}.
   In short, our main finding is that it looks like an explosion,
   with matter ``fireball'' expanding from the collision point, but a non-hydrodynamical
   explosion, in which fluid cannot be assigned temperature or entropy.
  
  What can be a physical significance and applications of these results? 
  
   Literally, they describe energy/momentum flow following a collision of say
   two heavy-quark mesons in a strongly coupled
    $\cal N$=4 gauge theory). It would then be instructive to compare
    these results with those in a weakly coupled regime of the same theory,
    in which the appropriate calculation would be perturbative radiation of massless
gluons and scalars. Those are well known to produce dipole radiation at small
velocities and bremmstrahlung cones (or ``jets'') in forward and backward
directions. In QCD perturbative radiation is affected by confinement
effects as well as the presence of light fermions: thus formation of QCD strings 
and their breaking by light quark pair production.
All of it is well modeled by QCD ``event generators'', one of which -- the Lund model --
we mentioned in the Introduction. 

So, why one would be interested in a strongly-coupled version of the ``event generator''?
One reason can be methodical: to better understand the difference between strongly coupled conformal
regime and confining theories, as far as jet physics is concerned. It has been
studied in literature that hypothetical  ``hidden valleys theories''
which may be found at LHC \cite{Han:2007ae} are strongly coupled:
so there is some nonzero (but tiny)  chance that those can be used in real 
experiments one day.

However, as explained in the Introduction, those were not our motivations.
We have done this calculation as a methodical step toward understanding
heavy ion collisions, in the AdS/CFT setting. One obviously needs to study one falling
string before considering many.  And simply adding the effect of many strings,
as we did above, is not yet sufficient to understand  heavy ion collisions.

As a discussion item, we would like at the end of the paper to indicate where
we will go from here. What we would like to understand in general is {\em how
and under which conditions the equilibration and entropy production happen,}
so that non-hydro explosion described above becomes hydro-like.
In order to derive that, one has to abandon the ``probe approximation''
used above, and include the gravitational impact of falling matter (strings in our setting)
in the metric. Only then one may see a transition from extremal black hole
(AdS metric we use) to non-extremal black hole with matter mass added
and a {\em nonzero horizon formed}. The horizon, when present, does provide both
Hawking temperature and Bekenstein entropy. We expect to use in the next
paper of this series a ``two-membrane paradigm'', in which collision debries
are represented
by one (falling) membrane, and the (rising and then falling because of stretching) horizon membrane by the other.
(These two membranes can be associated with the so called top-down and
down-up equilibration scenarios, proposed in literature in various model settings.)
When and where  most of the entropy is produced is the major issue to be addressed.  
How that is reflected in the ``hologram'' observed in the gauge theory could
then be calculated in the linearized approximation, as above.

\appendix
\section{Inverse Fourier Transform of Green's function}

Let us recall the expression of Green's function in momentum space:
\be
G(z,z')=\lbrace
\begin{array}{ll}
-{2\over{z'}}I_2(i\lam z_<)K_2(i\lam z>)& \omg>0,\lv\omg\rv>k {\I} \\
-{2\over{z'}}I_2(-i\lam z_<)K_2(-i\lam z>)& \omg<0,\lv\omg\rv>k {\II} \\
-{2\over{z'}}I_2(\tlam z_<)K_2(\tlam z>)& \lv\omg\rv<k {\III}
\end{array} \no
\ee

We will use $\I$,$\II$,$\III$ to refer to the three cases as indicated above.
In order to do the inverse Fourier transform: 
$\frac{1}{(2\pi)^4}\int G(z,z')e^{i\omg t-i\vk\vx}d\omg d^3k$, we make
the change of variable for each case:

\be\label{eq:omg}
&&{\I}:   \omg=\sqrt{k^2+\lam^2} \frac{1}{(2\pi)^4}\int G(z,z')e^{i\omg t-i\vk\vx}d\omg d^3k \no
&&=\frac{1}{(2\pi)^3}\int G(z,z')e^{i\sqrt{k^2+\lam^2}t}\frac{2sin(kr)}{kr}k^2dk\frac{\lam d\lam}{\sqrt{k^2+\lam^2}} \no
&&{\II}:  \omg=-\sqrt{k^2+\lam^2} \frac{1}{(2\pi)^4}\int G(z,z')e^{i\omg t-i\vk\vx}d\omg d^3k \no
&&=\frac{1}{(2\pi)^3}\int G(z,z')e^{-i\sqrt{k^2+\lam^2}t}\frac{2sin(kr)}{kr}k^2dk\frac{\lam d\lam}{\sqrt{k^2+\lam^2}} \no
&&{\III}:  \omg=\pm\sqrt{k^2-\tlam^2} \frac{1}{(2\pi)^4}\int G(z,z')e^{i\omg t-i\vk\vx}d\omg d^3k \no
&&=\frac{1}{(2\pi)^3}\int G(z,z')2cos{\sqrt{k^2-\tlam^2}t}\frac{2sin(kr)}{kr}k^2dk \no
&&\times\frac{\lam d\lam}{\sqrt{k^2-\tlam^2}}
\ee

We use the following formulas to evaluate the k-integrals:

\be\label{eq:cos}
&&2\int_0^\infty\frac{cos(b\sqrt{x^2+a^2})}{\sqrt{x^2+a^2}}sin(\xi x)xdx=
\pi Y_0'(a\sqrt{b^2-\xi^2})\times \no
&&\frac{-a\xi}{\sqrt{b^2-\xi^2}}\theta(b-\xi) 
-2K_0'(a\sqrt{\xi^2-b^2})\frac{a\xi}{\sqrt{\xi^2-b^2}}\theta(\xi-b) \no
&&2\int_0^\infty\frac{cos(b\sqrt{x^2-a^2})}{\sqrt{x^2-a^2}}sin(\xi x)xdx=
-2K_0'(a\sqrt{b^2-\xi^2})\times \no
&&\frac{-a\xi}{\sqrt{b^2-\xi^2}}\theta(b-\xi) 
+\pi Y_0'(a\sqrt{\xi^2-b^2})\frac{a\xi}{\sqrt{\xi^2-b^2}}\theta(\xi-b) \no
&&2\int_0^\infty\frac{sin(b\sqrt{x^2+a^2})}{\sqrt{x^2+a^2}}sin(\xi x)xdx=
-\pi J_0'(a\sqrt{b^2-\xi^2}) \no
&&\times\frac{-a\xi}{\sqrt{b^2-\xi^2}}\theta(b-\xi)+\pi\dlt(b-\xi)
\ee
with $a,b,\xi>0$

These formulas are obtained by differentiating with respect $\xi$
the cosine-transform of 
$\frac{cos(b\sqrt{x^2+a^2})}{\sqrt{x^2+a^2}}$,
$\frac{cos(b\sqrt{x^2-a^2})}{\sqrt{x^2-a^2}}$ and
$\frac{sin(b\sqrt{x^2+a^2})}{\sqrt{x^2+a^2}}$

Let's focus on the case $t>0$ at the moment. With the help of \ref{eq:cos},
\ref{eq:omg} becomes:

\be
&&{\I}: \int-\frac{2}{z'}I_2(i\lam z_<)K_2(i\lam z_>)\frac{\lam d\lam}{(2\pi)^3}\biggl[
\pi Y_0'(\lam\sqrt{t^2-r^2}) \no
&&\times\frac{-\lam}{\sqrt{t^2-r^2}}\theta(t-r) 
-2K_0'(\lam\sqrt{r^2-t^2})\frac{\lam}{\sqrt{r^2-t^2}}\theta(r-t) \no
&&-i\pi J_0'(\lam\sqrt{t^2-r^2})\frac{-\lam}{\sqrt{t^2-r^2}}\theta(t-r)
+i\pi\frac{\dlt(t-r)}{r}
\biggr] \no
&&{\II}: \int-2\frac{2}{z'}I_2(-i\lam z_<)K_2(-i\lam z_>)\frac{\lam d\lam}{(2\pi)^3}\biggl[
\pi Y_0'(\lam\sqrt{t^2-r^2}) \no
&&\times\frac{-\lam}{\sqrt{t^2-r^2}}\theta(t-r) 
-2K_0'(\lam\sqrt{r^2-t^2})\frac{\lam}{\sqrt{r^2-t^2}}\theta(r-t) \no
&&+i\pi J_0'(\lam\sqrt{t^2-r^2})\frac{-\lam}{\sqrt{t^2-r^2}}\theta(t-r)
-i\pi\frac{\dlt(t-r)}{r}
\biggr] \no
&&{\III}: 2\int-\frac{2}{z'}I_2(\lam z_<)K_2(\lam z_>)\frac{\lam d\lam}{(2\pi)^3}
\biggl[-2K_0'(\lam\sqrt{t^2-r^2}) \no
&&\times\frac{-\lam}{\sqrt{t^2-r^2}}\theta(t-r)
+\pi Y_0'(\lam\sqrt{r^2-t^2}\frac{\lam}{\sqrt{r^2-t^2}}\theta(r-t))
\biggr] \no
\ee

Suppose we replace $z_>$ by $z_>-i\epsilon$, convergence of the integral
enables us to rotate the contour of ${\III}$ and ${\I}$:
\be
&&\int-\frac{2}{z'}I_2(\lam z_<)K_2(\lam z_>)\frac{\lam d\lam}{(2\pi)^3}
\biggl[-2K_0'(\lam\sqrt{t^2-r^2}) \no
&&\times\frac{-\lam}{\sqrt{t^2-r^2}}\theta(t-r)
\biggr]
=\int-\frac{2}{z'}I_2(i\lam z_<)K_2(i\lam z_>)\frac{\lam d\lam}{(2\pi)^3}\times \no
&&\biggl[(-\pi Y_0'(\lam\sqrt{t^2-r^2})-i\pi J_0'(\lam\sqrt{t^2-r^2}))
\frac{-\lam}{\sqrt{t^2-r^2}}\theta(t-r)
\biggr] \no
&&\int-\frac{2}{z'}I_2(i\lam z_<)K_2(i\lam z_>)\frac{\lam d\lam}{(2\pi)^3}
\biggl[-2K_0'(\lam\sqrt{r^2-t^2}) \no
&&\times\frac{\lam}{\sqrt{r^2-t^2}}\theta(r-t)
\biggr]
=\int-\frac{2}{z'}I_2(\lam z_<)K_2(\lam z_>)\frac{\lam d\lam}{(2\pi)^3}\times \no
&&\biggl[(-\pi Y_0'(\lam\sqrt{r^2-t^2})+i\pi J_0'(\lam\sqrt{r^2-t^2}))
\frac{\lam}{\sqrt{r^2-t^2}}\theta(r-t) \biggr] \no
\ee

Similarly, suppose $z_>\rightarrow z_>+i\epsilon$, we can rotate the contour
of ${\III}$ and ${\II}$:
\be
&&\int-\frac{2}{z'}I_2(\lam z_<)K_2(\lam z_>)\frac{\lam d\lam}{(2\pi)^3}
\biggl[-2K_0'(\lam\sqrt{t^2-r^2})\times \no
&&\frac{-\lam}{\sqrt{t^2-r^2}}\theta(t-r)
\biggr]
=\int-\frac{2}{z'}I_2(-i\lam z_<)K_2(-i\lam z_>)\frac{\lam d\lam}{(2\pi)^3}\times \no
&&\biggl[(-\pi Y_0'(\lam\sqrt{t^2-r^2})+i\pi J_0'(\lam\sqrt{t^2-r^2}))
\frac{-\lam}{\sqrt{t^2-r^2}}\theta(t-r)
\biggr] \no
&&\int-\frac{2}{z'}I_2(-i\lam z_<)K_2(-i\lam z_>)\frac{\lam d\lam}{(2\pi)^3}
\biggl[-2K_0'(\lam\sqrt{r^2-t^2})\times \no
&&\frac{\lam}{\sqrt{r^2-t^2}}\theta(r-t)
\biggr]
=\int-\frac{2}{z'}I_2(\lam z_<)K_2(\lam z_>)\frac{\lam d\lam}{(2\pi)^3}\times \no
&&\biggl[(-\pi Y_0'(\lam\sqrt{r^2-t^2})-i\pi J_0'(\lam\sqrt{r^2-t^2}))
\frac{\lam}{\sqrt{r^2-t^2}}\theta(r-t) \biggr] \no
\ee

Summing up piece ${\I}$,${\II}$ and ${\III}$, we find various terms cancel against each other.
We are left with:

\be
&&z_>\rightarrow z_>-i\epsilon: \no
&&\int-\frac{2}{z'}I_2(i\lam z_<)K_2(i\lam z_>)\frac{\lam d\lam}{(2\pi)^3}
\biggl[-2\pi iJ_0'(\lam\sqrt{t^2-r^2}) \no
&&\times\frac{-\lam}{\sqrt{t^2-r^2}}\theta(t-r)
+\pi i\frac{\dlt(t-r)}{r} \biggr]
+\int-\frac{2}{z'}I_2(\lam z_<)K_2(\lam z_>) \no
&&\times\frac{\lam d\lam}{(2\pi)^3}
\biggl[\pi iJ_0'(\lam\sqrt{r^2-t^2})\frac{\lam}{\sqrt{r^2-t^2}}\theta(r-t)
\biggr] \no
&&z_>\rightarrow z_>+i\epsilon: \no
&&\int-\frac{2}{z'}I_2(-i\lam z_<)K_2(-i\lam z_>)\frac{\lam d\lam}{(2\pi)^3}
\biggl[2\pi iJ_0'(\lam\sqrt{t^2-r^2}) \no
&&\times\frac{-\lam}{\sqrt{t^2-r^2}}\theta(t-r)
-\pi i\frac{\dlt(t-r)}{r} \biggr]
+\int-\frac{2}{z'}I_2(\lam z_<)K_2(\lam z_>) \no
&&\times\frac{\lam d\lam}{(2\pi)^3}
\biggl[-\pi iJ_0'(\lam\sqrt{r^2-t^2})\frac{\lam}{\sqrt{r^2-t^2}}\theta(r-t)
\biggr] \no
\ee

If we are only interested in the coefficient of the $z_<^2$ term\footnote
{By analyzing the small $z$ expansion of $h_{mn}(z)=\int G(z,z')s_{mn}(z')
dz'd^4x$, one can show the coefficient of $z^2$ term equals
that of $z_<^2$ term, provided that $s_{mn}(z')$ contains no $z'^0$ and
$z'^2$ terms. The latter condition is satisfied by the sources considered
in this paper}, we may make the following substitution: 
$I_2(\pm i\lam z_<)\rightarrow-\frac{1}{8}\lam^2$,$I_2(\lam z_<)\rightarrow \frac{1}{8}\lam^2$,$z_>\rightarrow z'$

Further evaluation of the integral involves the two formulas:

\be
\int_0^\infty x^{\mu+\nu+1}J_{\mu}(ax)K_{\nu}(bx)dx=2^{\mu+\nu}a^{\mu}b^{\nu}
\frac{\Gamma(\mu+\nu+1)}{(a^2+b^2)^{\mu+\nu+1}} \no
\int_0^\infty x^{\mu}K_{\nu}(ax)dx=2^{\mu-1}a^{-\mu-1}
\Gamma(\frac{1+\mu+\nu}{2})\Gamma(\frac{1+\mu-\nu}{2}) \no
\ee

It is easy to find $\int\lam^3 d\lam K_2(\pm i\lam z_>)$ and 
$\int\lam^4 d\lam K_2(\lam z_>)J_1(\lam\sqrt{r^2-t^2})\theta(r-t)$ contain
no singularity in $z_>$, thus the contribution from $z_>\pm i\epsilon$ cancel
each other. We end up with:

\be
&&\int-\frac{2}{z'}\frac{-\lam^2}{8}K_2(i\lam z_>)\frac{\lam d\lam}{(2\pi)^3}
(-2\pi i)J_1(\lam\sqrt{t^2-r^2}) \no
&&\times\frac{\lam}{\sqrt{t^2-r^2}}\theta(t-r)
\Big\vert_{z_>\rightarrow z_>-i\epsilon}
+\int-\frac{2}{z'}\frac{-\lam^2}{8}K_2(-i\lam z_>) \no
&&\times\frac{\lam d\lam}{(2\pi)^3}
2\pi iJ_1(\lam\sqrt{t^2-r^2})\frac{\lam}{\sqrt{t^2-r^2}}\theta(t-r)
\Big\vert_{z_>\rightarrow z_>+i\epsilon} \no
&&=\frac{12iz'}{(2\pi)^2}
\biggl[\frac{1}{(t^2-r^2-z'^2+i\epsilon)^4}-\frac{1}{(t^2-r^2-z'^2-i\epsilon)^4}\biggr] \no
&&\times\theta(t-r)
\ee

For $t<0$, similar procedure leads to a vanishing result. Therefore the
 $z_<^2$ term of the Green's function in coordinate space, which is
exactly the propagator we are looking for, can be expressed as
\footnote{the $t>0$ condition is included in the theta function}:

\be
&&P_R=\frac{12iz'}{(2\pi)^2}
\biggl[\frac{1}{(t^2-r^2-z'^2+i\epsilon)^4}- \no
&&\frac{1}{(t^2-r^2-z'^2-i\epsilon)^4}\biggr]\theta(t-r)
\ee

\section{Extract Energy Density in Comoving Frame}

The aim is to kill the $(0,i)$ components(momentum density)
 by local Lorentz transformation: $T'=STS^T$, where
\be
T'=T'_{\alpha\beta}\;,T=T_{\mu\nu}\;,S_{\alpha\mu}
=\frac{\del x^{\mu}}{\del x^{\alpha}}
\ee

Local Lorentz transformation matrix is a product of matrices, which 
are either rotations,e.g.$\I$ or boosts,e.g.$\II$.
This is a consequence of the fact: $SgS^T=g$, where $g=diag(-1,1,1,1)$ is the
Minkowski metric.
\be
&&\I
\begin{pmatrix}
1& & & \\
& \cos(\theta)& \sin(\theta)& \\
& -\sin(\theta)& \cos(\theta)& \\
& & & 1
\end{pmatrix} \no
&&\II
\begin{pmatrix}
chy& shy& & \\
shy& chy& & \\
& & 1& \\
& & & 1
\end{pmatrix} \no
\ee

A nice property of $g$ is: $g=g^T=g^{-1}$, therefore $gT'$ and $Tg$ 
are related by similarity transformation:

\be
gT'=gSTgg^TS^T=gS(Tg)(gS)^{-1}
\ee

Since $gT'$ does not alter the zero entries of $T'$, now the problem
becomes to kill the $(0,i)$ components of $Tg$ by similarity transformation.
The original matrix $Tg$ can be
viewed as an operator $L$ acting on a set of basis, while the similarity
transformation is just a change of basis:
\be
&&(Tg)_{mn}=(e_n,Le_m) \no
&&(gT')_{mn}=(e_n',Le_m') \no
&&e_n'=(gS)_{nm}e_m
\ee

We want to find a basis $e_0'$ such that $Le_0'=\lam e_0'$ with $\lam=(gT')_{00}$. Denote $e_0'=x_m e_m$, it is easy to show $x_m(Tg)_{mn}=\lam x_n$
This is exactly an eigenvalue problem for matrix $(Tg)^T$. The restriction in
transformation matrix is translated to: $x_0^2-x_1^2-x_2^2-x_3^2>0$. It turns
out this condition is just enough to determine a unique eigenvalue out of
four possible eigenvalues. The energy density is given by $\epsilon=-\lam$.

\section*{Acknowledgment}

This work is  supported by the US-DOE grants DE-FG02-88ER40388
and DE-FG03-97ER4014.


\begin{thebibliography}{99}
\bibitem{Lin:2006rf}
  S.~Lin and E.~Shuryak,
  arXiv:hep-ph/0610168.


\bibitem{MLV}
  L.~D.~McLerran and R.~Venugopalan,
  Phys.\ Rev.\  D {\bf 49}, 2233 (1994)
  [arXiv:hep-ph/9309289].


\bibitem{Andersson:1983ia}
  B.~Andersson, G.~Gustafson, G.~Ingelman and T.~Sjostrand,
  Phys.\ Rept.\  {\bf 97}, 31 (1983).

\bibitem{Brodsky:2007vk}
  S.~J.~Brodsky and G.~F.~de Teramond,
  arXiv:0709.2072 [hep-ph].


\bibitem{Polchinski:2001tt}
  J.~Polchinski and M.~J.~Strassler,
  Phys.\ Rev.\ Lett.\  {\bf 88}, 031601 (2002)
  [arXiv:hep-th/0109174].

\bibitem{mydomain}E.Shuryak, 
   A "Domain Wall" scenario for the AdS/QCD,arXiv:0711.0004

\bibitem{Giddings:2000ay}
  S.~B.~Giddings and E.~Katz,
  J.\ Math.\ Phys.\  {\bf 42}, 3082 (2001)
  [arXiv:hep-th/0009176].

\bibitem{Lin:2007pv}
  S.~Lin and E.~Shuryak,
  Phys.\ Rev.\  D {\bf 76}, 085014 (2007)
  [arXiv:0707.3135 [hep-th]].

\bibitem{Gubser} J.~J. Friess, S.~S. Gubser, G.~Michalogiorgakis
JHEP09, 072 (2006)
  {{\tt hep-th/0605292}}.

\bibitem{Chesler:2007an}
  P.~M.~Chesler and L.~G.~Yaffe,
  arXiv:0706.0368 [hep-th].

\bibitem{son:correlator}
  D.~T.~Son and A.~O.~Starinets,
  JHEP {\bf 0209}, 042 (2002)
  [arXiv:hep-th/0205051].

\bibitem{vacua_probes}
  U.~H.~Danielsson, E.~Keski-Vakkuri and M.~Kruczenski,
  JHEP {\bf 9901}, 002 (1999)
  [arXiv:hep-th/9812007].

\bibitem{Giddings:2000mu}
  S.~B.~Giddings, E.~Katz and L.~Randall,
  JHEP {\bf 0003}, 023 (2000)
  [arXiv:hep-th/0002091].

\bibitem{shock}
  G.~T.~Horowitz and N.~Itzhaki,
  JHEP {\bf 9902}, 010 (1999)
  [arXiv:hep-th/9901012].

\bibitem{stone}
  J.~J.~Friess, S.~S.~Gubser, G.~Michalogiorgakis and S.~S.~Pufu,
  JHEP {\bf 0704}, 080 (2007)
  [arXiv:hep-th/0611005].

\bibitem{KMT}
  I.~R.~Klebanov, J.~M.~Maldacena and C.~B.~.~Thorn,
  JHEP {\bf 0604}, 024 (2006)
  [arXiv:hep-th/0602255].

\bibitem{Teaney:2001av}
  D.~Teaney, J.~Lauret and E.~V.~Shuryak,
  arXiv:nucl-th/0110037.

\bibitem{Kolb:2003dz}
  P.~F.~Kolb and U.~W.~Heinz,
  arXiv:nucl-th/0305084.

\bibitem{Han:2007ae}
  T.~Han, Z.~Si, K.~M.~Zurek and M.~J.~Strassler,
  arXiv:0712.2041 [hep-ph].

\end{thebibliography}
\end{document}